\documentclass{article}
\usepackage[super,sort&compress,comma]{natbib}
\usepackage{url}
\usepackage{graphicx}
\usepackage{bookmark}
\usepackage{longtable}
\usepackage{rotating} 
\usepackage{hyperref}
\usepackage{amsmath}
\usepackage{amsfonts}
\usepackage{booktabs}
\usepackage{colortbl}
\usepackage{authblk}
\usepackage[font=scriptsize,labelfont=bf]{caption}
\usepackage{geometry}
\begin{document}

\title{Artificial Intelligence and Generative Models for Materials Discovery: A Review }

\author[1]{Albertus Denny Handoko}
\affil[1]{Institute of Sustainability for Chemicals, Energy and Environment (ISCE$^2$)\\
Agency for Science, Technology and Research (A*STAR)\\
1 Pesek Road, Jurong Island, Singapore 627833, Republic of Singapore\\
handoko\_albertus@isce2.a-star.edu.sg\\handoko\_albertus@a-star.edu.sg}

\author[2]{Riko I Made\footnote{Corresponding author:riko@a-star.edu.sg}}
\affil[2]{Institute of Materials Research and Engineering (IMRE)\\
Agency for Science, Technology and Research (A*STAR)\\
2 Fusionopolis Way, Innovis \#08-03, Singapore 138634, Republic of Singapore\\
riko@imre.a-star.edu.sg\\riko@a-star.edu.sg\\
}

\date{} 
\maketitle

\begin{abstract}
High throughput experimentation tools, machine learning (ML) methods, and open material databases are radically changing the way new materials are discovered. From the experimentally driven approach in the past, we are moving quickly towards the artificial intelligence (AI) driven approach, realising the 'inverse design' capabilities that allow the discovery of new materials given the desired properties. 
This review aims to discuss different principles of AI-driven generative models that are applicable for materials discovery, including different materials representations available for this purpose. We will also highlight specific applications of generative models in designing new catalysts, semiconductors, polymers, or crystals while addressing challenges such as data scarcity, computational cost, interpretability, synthesizability, and dataset biases. Emerging approaches to overcome limitations and integrate AI with experimental workflows will be discussed, including multimodal models, physics-informed architectures, and closed-loop discovery systems. This review aims to provide insights for researchers aiming to harness AI’s transformative potential in accelerating materials discovery for sustainability, healthcare, and energy innovation.  
\end{abstract}


\section{Introduction}
\label{sec:intro}

\begin{figure}[bt]
    
    \centering
    \includegraphics[width=0.80\linewidth]{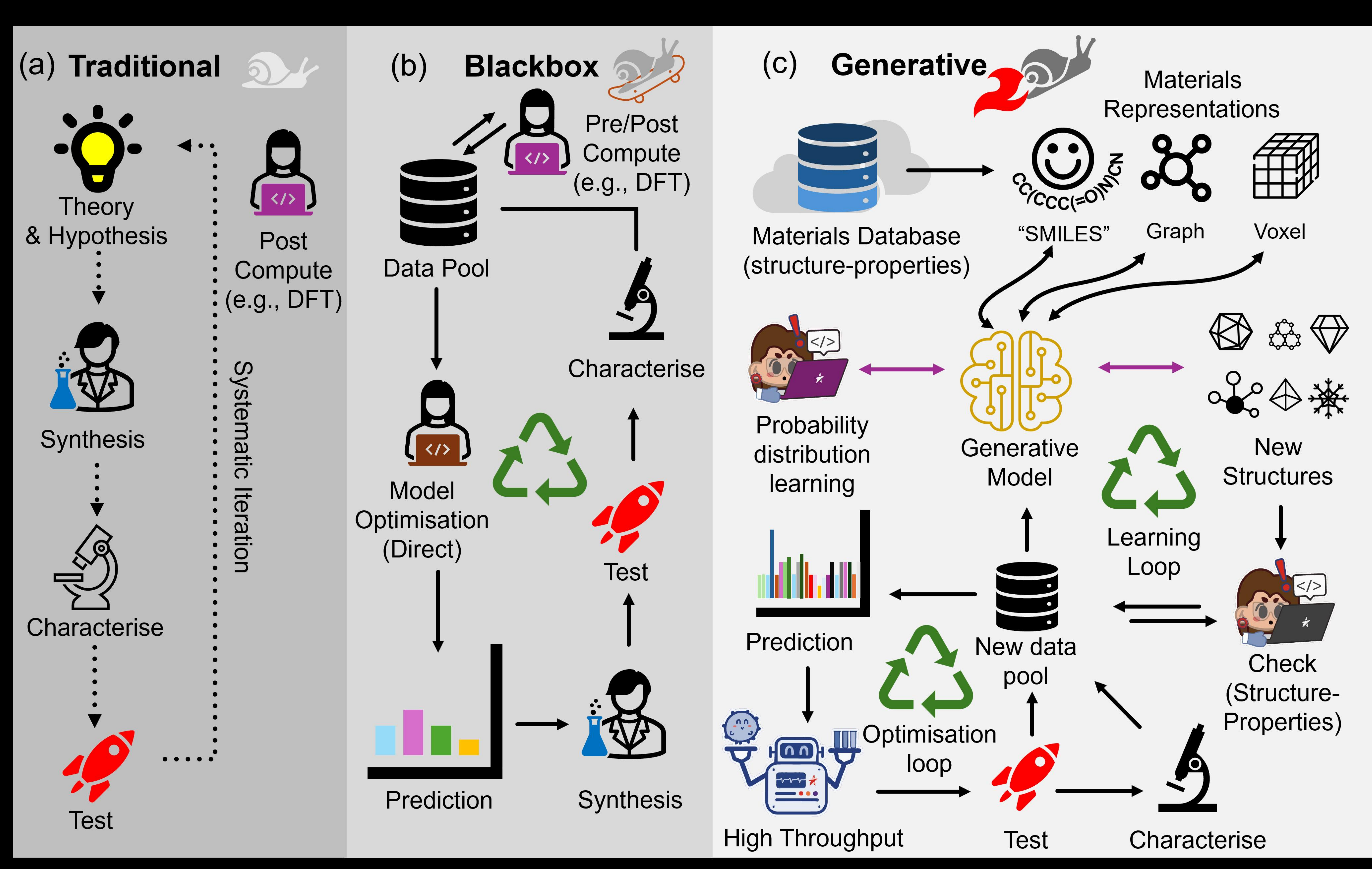}
    \scriptsize
    \caption{A paradigm shift in Materials Discovery. The strategy for discovering new materials has evolved over the last century. (a) Traditionally, materials discovery starts with an idea or hypothesis that needs to be validated through experimental synthesis, characterization, and testing. The advent of computational techniques allows \textit{ab-initio} evaluation of the novel materials to gain deeper insights into the structure-properties relationship or guide subsequent discoveries. (b) Today, we witness many "blackbox" discovery approaches that allow an iterative direct model optimization process. With more accessible and efficient computational resources, computational techniques, including \textit{ab-initio} calculations, can be implemented in the pre-optimisation step, where data pools can be enriched by computational inputs. A second, more thorough \textit{ab-initio} calculation may be performed on the optimum material to validate or further elaborate the material-structure relationship. (c) In the future, we predict a growing "Generative Discovery" approach that leverages existing materials databases rich in past learnings. Ways to encode materials structure into computer-readable representations like "SMILES" \cite{weiningerSMILESChemicalLanguage1988}, graph\cite{RN7143_Graph}, or voxel\cite{RN7138_Orlando_Voxel} allow effective exploration of the materials space.  Unlike previous approaches, the generative approach can learn probability distributions, capable of suggesting innovative material structures even before the experiments begin. As the experiment progresses, the continuous learning loop refines the model with fresh data, while predicted structures are meticulously evaluated against target properties, followed by high-throughput synthesis, characterization, and testing. CC(CCC(=O)N)CN is the SMILES representation of 5-amino-4-methylpentanamide }
    \label{fig:discovery}
\end{figure}

Materials science is the foundation for technological innovation, driving advances in energy, electronics, catalysis, and quantum computing through the development of novel materials with tailored properties \citep{butlerMachineLearningMolecular2018, vasudevanMaterialsScienceArtificial2019}. Historically, material discovery is experiment-driven. Often, this means labourious trial and error process where scientists first hypothesize the structures, synthesize compounds, and then test properties \citep{correa-baenaAcceleratingMaterialsDevelopment2018} (Fig. \ref{fig:discovery}a).  While there is nothing fundamentally wrong with this approach \cite{RN7021}, the vastness of chemical space, estimated to exceed $10^{60}$ carbon-based molecules, renders exhaustive experiment-led exploration to find new classes of materials impractical \citep{dobsonChemicalSpaceBiology2004, schneiderComputerbasedNovoDesign2005}. Consequently, the timeline from material conception to deployment often spans decades, hindering innovation and investment \citep{correa-baenaAcceleratingMaterialsDevelopment2018, vasylenkoElementSelectionCrystalline2021}.
Modern technologies such as electric vehicles, high-speed rails, and satellite communications demand new materials with lower weights and enhanced properties, such as high thermal conductivity, electromagnetic shielding, customisable bandgap, or enhanced mechanical strength, pushing the limits of existing compounds \citep{nohInverseDesignSolidState2019, zhangMachineLearningAssisted2021}. This bottleneck spurs the search for a novel approach to the discovery of materials that are capable of navigating complex structural and functional requirements. 

The most popular approach in harvesting the low-hanging fruits in the exploration of the vast materials space is what we call the "black-box" approach (Fig. ~\ref{fig:discovery}b). This is a general term representing intelligent data acquisition strategies that utilize ML optimization algorithms to find the right candidate(s) with targeted properties \cite{RN7131_Chitturi}. A black-box approach typically involves building a new dataset, built specifically for a particular optimization of a desired output (e.g., material properties) based on empirical parameter inputs (e.g., precursor ratio). This approach has been proven effective for well-defined and constrained problems, such as finding the appropriate precursors to optimize the catalytic condition \cite{RN5345_Lim}, or to optimise process parameters of a reaction \cite{RN5650_Gupta}\cite{RN6997_Mehta} in which much fewer iteration steps are required compared to traditional methods. However, it is difficult to generalize a black box approach that has been trained in a specific task, unless there are some similarities in properties or structure in the related task \cite{RN7134_Jiang}. 

Recognising the limitations of the blackbox approaches to discover truly new materials that can display groundbreaking properties, multi-disciplinary scientists started to apply generative models for materials discovery. Originally designed as AI to simulate human reasoning, intelligence, and creative processes \cite{RN7135_Rosenblatt}, generative model are gradually finding their way to materials discovery applications. Generative models are not adopted into materials discovery overnight. We recognise at least five key "parents" that brought us AI-driven materials discovery (Fig.~\ref{fig:discovery}c). 

The first parent is not actually related to AI (or ML), but rather high-throughput combinatorial methods and tools development. Combinatorial methods have been ubiquitous in nature. For example, to obtain suitable antibodies to fight certain pathogens, lymphocytes are assembled in the human body by recombination of  large "libraries" of molecules and selecting those with desired properties or mutating them \cite{RN7014}. Such an approach is only relatively recently being applied to materials science \cite{RN7011,RN7013} , where large arrays of materials composition are being synthesised (for example, by inkjet\cite{RN7136_Zeng} or plasma printing\cite{RN6094_agrotis}) for subsequent systematic testing. Today, coordinated efforts, with many research centres focusing on high throughput experimentation worldwide. 

The second parent is the application of ML algorithms for parametric optimisation. The advent of machine learning (ML) has revolutionized materials science by leveraging vast datasets and computational power to uncover intricate patterns and accelerate discovery \citep{butlerMachineLearningMolecular2018, merchantScalingDeepLearning2023, zuoAcceleratingMaterialsDiscovery2021}.  The syntheses of organic and inorganic materials can be complicated and challenging to optimise, as they often involve multiple steps and precursors. Whilst existing statistical optimisation approaches like the design of experiment (DOE) methods have proven to be instrumental in the discovery of new materials \cite{RN7029Gisperg}, the integration of ML can enhance the efficiency and efficacy of these methods further, especially for more complex materials composition or metastable compounds. The synergy between ML, domain expertise, and established DOE methods has been observed to be more robust compared to either standalone approaches \cite{RN7027Rummukainen}. 

The third is the sharing of materials databases\cite{FAIRPrinciples}. The mountains of data generated by high-throughput combinatorial methods are only beneficial if it is shared among the wider scientific community. Researchers around the world recognises this, and there has been an exponential increase in curated materials databases in many countries  \cite{RN7023,RN7024,RN7030,RN2007MaterialsProject}. However, differences in the way labs around the world perform the experiments and record the findings can give rise to dataset mismatch or variation. Developing universally accepted standards is also challenging, as the high-throughput combinatorial methods are still in the infancy with rapidly changing protocols, tools, and algorithms. We note two major efforts to develop standardised testing and recording computerised materials data have been performed by the Versailles Project on Advanced Materials and Standards (VAMAS, technical work area 10)\cite{VersaillesProjectAdvanced} and ASTM International (Committee E-49) \citep{RN7015Reynard}, although implementation of these standards in research laboratories remain scarce. 

The fourth is the application of ML in the computational modeling of the force field\cite{RN6303Unke}. This is a significant advancement in computational chemistry that bridges between very precise atomic modeling through \textit{ab initio} density functional theory (DFT) calculation and the force field model that drives molecular dynamics (MD). The two computational approaches are on different ends of scale and accuracy: DFT glances from the quantum mechanical point of view, capable of calculating accurate atomic interaction and potential of considered systems, but can only cover a limited (angstroms) range due to the rigorous calculation steps. On the other hand, MD approximates atoms and molecules as particles and uses simpler classical mechanics to solve the dynamic behaviour of a larger number of atoms over a (brief) period. Essential to MD's capability to simulate the dynamic behaviour is the force field model \cite{RN7037_ForceField}, an empirical method to describe the interactions between atoms in the system without the need to model the entire electronic structure or interatomic potential. The implementation of ML methods, especially machine-learned potential (MLP)\cite{RN7043_Friederich_MLPotential}, allows a dream of "hybrid" simulation where accurate potential energies of a larger system (or over a longer period) can be quickly obtained from a suitable numerical representation of the material \cite{RN7044_Collins_MLPDescriptor} that can typically be trained with a pool of ab-initio simulation data or experimental data \cite{RN7018_Rocken_MLPFusion}.

Last but not least, is the incorporation of generative models into materials discovery. Generative models are able to approximate high-dimensional probability distributions between structures and desired characteristics or properties \citep{gomez-bombarelliAutomaticChemicalDesign2018, ruthottoIntroductionDeepGenerative2021, anstineGenerativeModelsEmerging2023, sanchez-lengelingInverseMolecularDesign2018}. Once the probability distributions have been learned, novel data such as molecular structures can be generated by sampling in the probability distributions' (latent) space, based on, for example, the desired properties. The ability of generative models to generate new structure suggestions from the latent space represents a new paradigm of materials discovery. This is a marked departure from previous approaches, where new structure suggestions need to be first explicitly generated in the real space, either by modifying or substituting the atoms found in known structures \cite{RN7050_Hautier_Substitution}, or by placing completely random atoms within preselected constraints or restraints to ensure that the generated materials are stable and unique \cite{RN7046_Pickard_Randomstructure}.  The next few sections of this review will focus on this fifth "parent", describing the different models available for materials discovery and their principles, along with examples and applications of these models in research.  

\section{Generative Models for Materials Science}

\begin{figure}[bt]
    \centering
    \includegraphics[width=0.80\linewidth]{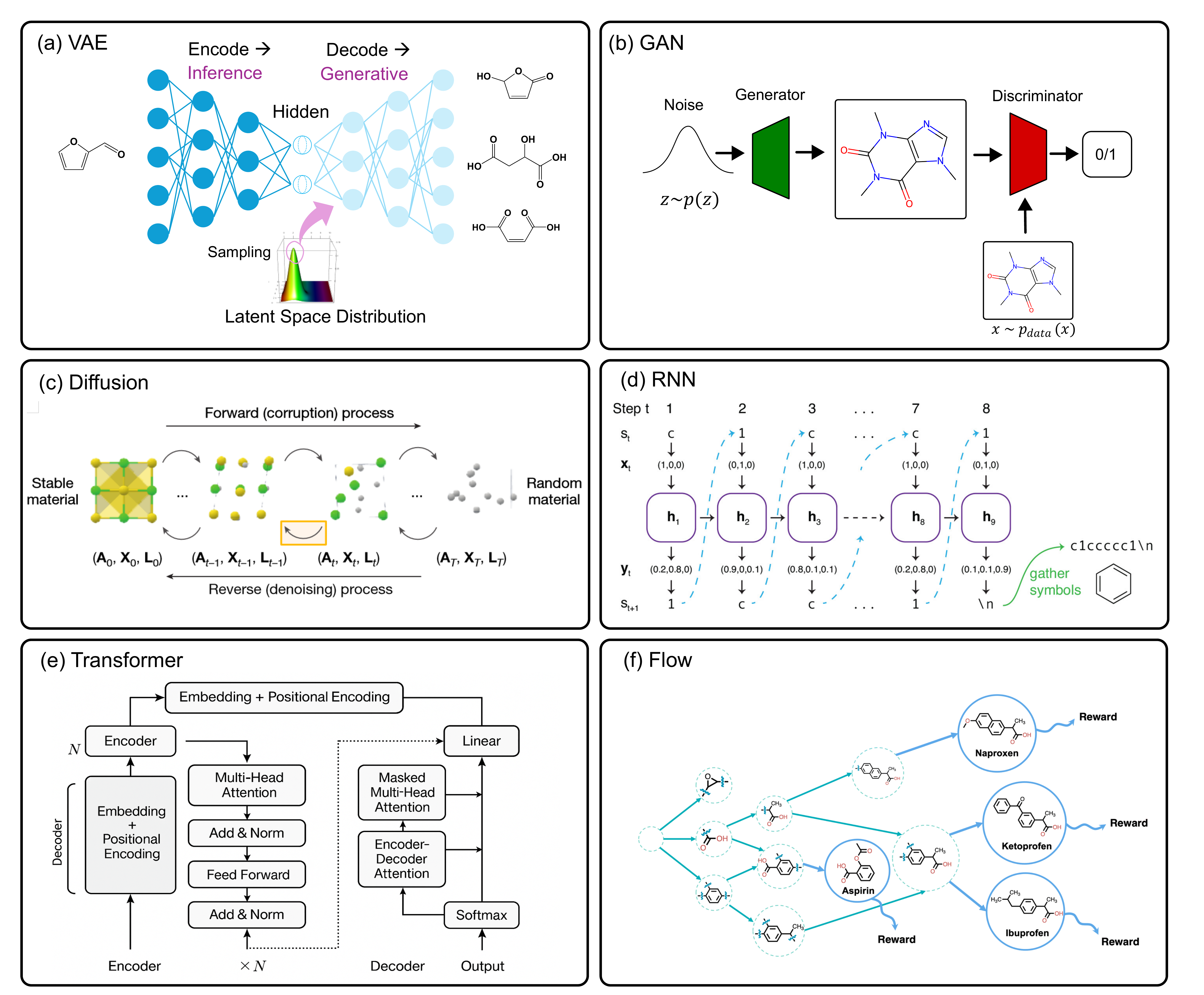}
    \scriptsize
    \caption{Schematics of generative model architectures for materials discovery, illustrating the general workflows for  (a) \textbf{VAE (Variational Autoencoder):} encode-decode process with an inference and generative path, showcasing the mapping of molecules into a latent space distribution and subsequent generation of new molecular structures. (b) \textbf{GAN (Generative Adversarial Network):}  a generator creating molecular structures from noise, which are then evaluated by a discriminator against real molecular data to improve realism. (c) \textbf{Diffusion Model:}  a forward (corruption) process where stable material is progressively turned into random material, and a reverse (denoising) process that reconstructs the stable material. This is analogous to generating molecules by reversing a corruption process. Taken with permission from \cite{alakhdarDiffusionModelsNovo2024}. (d) \textbf{RNN (Recurrent Neural Network):}  A sequential process where hidden states ($h_t$) are updated based on previous states and current inputs ($x_t$), eventually leading to the gathering of symbols (e.g., for molecular string generation). Taken with permission from  \cite{seglerGeneratingFocusedMolecule2018}. (e) \textbf{Transformer:}  encoder-decoder architecture with embedding, positional encoding, multi-head attention mechanisms, and feed-forward layers, commonly used for sequence-to-sequence tasks like molecular string manipulations,  (f) \textbf{Flow (Reinforcement Learning/Generative Flow):}  tree-like structure where molecular syntheses or modifications (e.g., from Aspirin to Naproxen, Ketoprofen, or Ibuprofen) are associated with rewards, suggesting a reinforcement learning approach for optimizing molecular properties or synthesis pathways. Taken with permission from \cite{jainGFlowNetsAIdrivenScientific2023}.}
    \label{fig:generative_models}
\end{figure}

To understand the role of generative models in materials discovery, it is essential to distinguish them from supervised machine learning paradigms. Supervised learning focuses on learning a mapping function, \(y = f(x)\), to predict outputs \(y\) from inputs \(x\) using labeled data, minimizing discrepancies between predicted and actual outcomes. Termed \textit{discriminative} models \citep{anstineGenerativeModelsEmerging2023}, these approaches excel in classification and regression tasks but are limited by their reliance on labeled datasets. In materials discovery, where novel structures and properties are often sought without extensively labeled data, generative models offer a powerful alternative.

Unlike discriminative models, generative models learn the underlying probability distribution, \(P(x)\), of the data, enabling the creation of new samples that closely resemble the training set. By capturing the inherent patterns in materials representations (more on this in section 2.2), these models can generate synthetic instances, often in unsupervised settings, leveraging both labeled and unlabeled data. A critical feature of the generative model is the \textit{latent space}: a low(er)-dimensional representation of the structure-properties relationship that enables inverse design strategy.

To understand how inverse design is achieved through generative models, six key types of generative models will be explored. These models are selected for their diverse principles and proven effectiveness in inverse design \textemdash  generating stable and novel materials for applications like catalysts, electronics, and polymers. We will discuss them in the order of historical emergence and increasing specialization, starting from Variational Autoencoders (VAEs), Generative Adversarial Networks (GANs), Diffusion Models (e.g., DiffCSP \citep{jiaoCrystalStructurePrediction2023}, SymmCD \citep{levySymmCDSymmetryPreservingCrystal2025}), Recurrent Neural Networks (RNNs) and Transformers (e.g., MatterGPT \citep{chenMatterGPTGenerativeTransformer2024}, Space Group Informed Transformer \citep{caoSpaceGroupInformed2024}), Normalizing Flows (e.g., CrystalFlow \citep{luoCrystalFlowFlowBasedGenerative2025}, FlowLLM \citep{sriramFlowLLMFlowMatching2024}), and Generative Flow Networks (GFlowNets, e.g., Crystal-GFN \citep{ai4scienceCrystalGFNSamplingCrystals2023})(Fig.~\ref{fig:generative_models}). 

The success of these models depends on effective material representations that preserve structural constraints, atomic interactions, and scalability across small molecules to large crystalline systems. Representations such as sequence-based (e.g., SMILES \citep{weiningerSMILESChemicalLanguage1988}), graph-based \citep{xieCrystalGraphConvolutional2018}, voxel-based, and physics-informed formats \citep{musilPhysicsInspiredStructuralRepresentations2021} enable models to handle complex materials data. By integrating these representations, generative models, from VAEs to GFlowNets, address diverse challenges in materials discovery, offering scalable solutions for crystalline, polymeric, and composite systems (Fig.~\ref{fig:representations}).

\subsection{Models and Principles}

\subsubsection{Variational Autoencoders (VAEs)}

\begin{figure}[bt]
    \centering
    \includegraphics[width=0.50\linewidth]{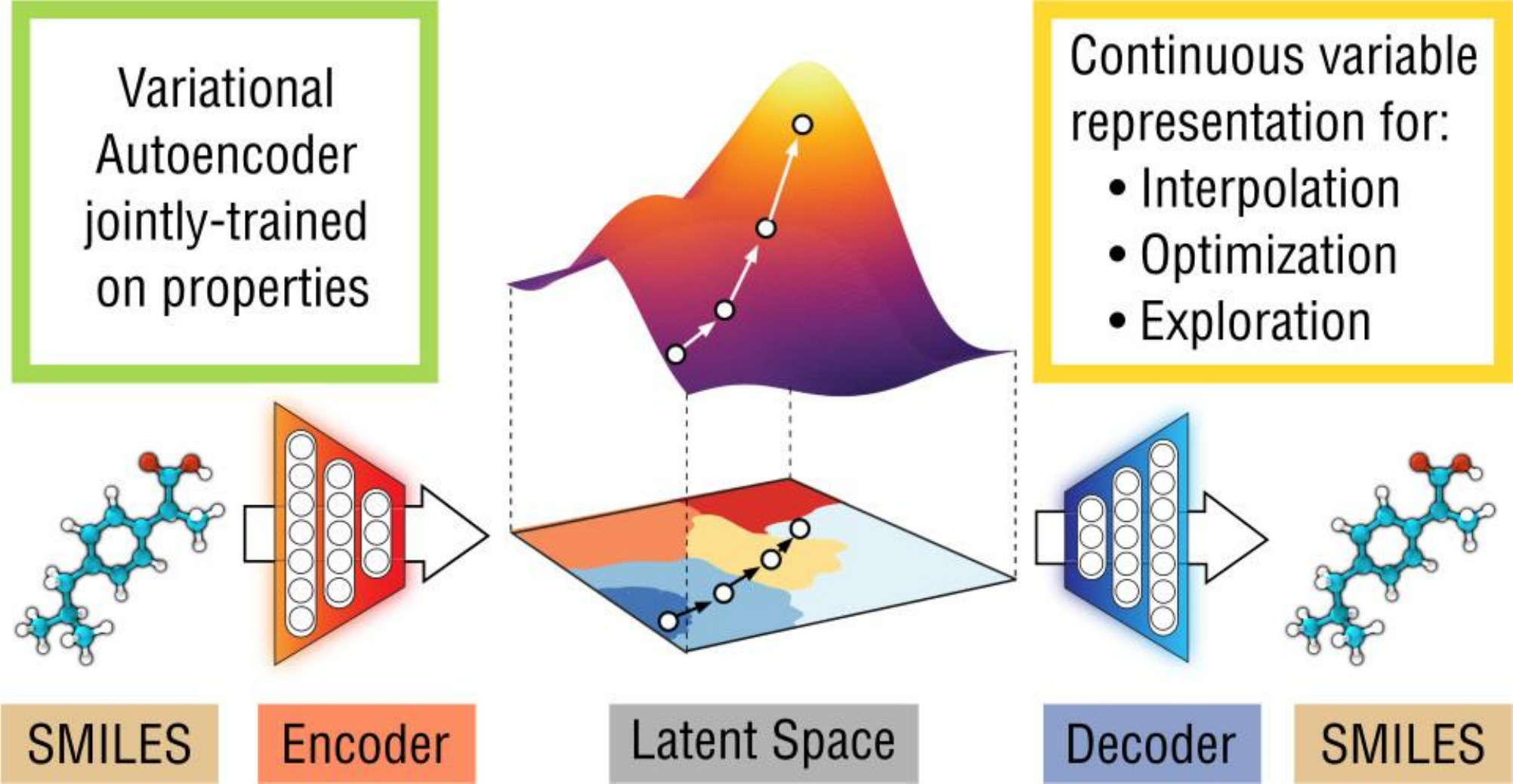} 
    \scriptsize
    \caption{An illustration of the use of VAE  for molecular design,  integrated with a joint property prediction model \textemdash \emph{predictor}. The \textbf{encoder} transforms discrete molecular representations (like SMILES strings) into continuous latent space vectors. The \textbf{decoder} then converts these latent vectors back into SMILES strings. The \emph{predictor} can be added to predict properties from latent representations. However, the huge size of the latent space dimension (more than 100) makes  sampling and visualization difficult. Taken with permission from \citet{gomez-bombarelliAutomaticChemicalDesign2018} }
    \label{fig:gomez2018_fabs}\end{figure}

Variational Autoencoders (VAEs) are generative models that learn a probabilistic latent space for data generation \citep{kingmaAutoEncodingVariationalBayes2022}. VAE typically consists of an encoder that maps input data $x$ (e.g., material descriptors) to a latent distribution $q(z|x) = \mathcal{N}(\mu(x), \sigma(x)^2)$, and a decoder that reconstructs $x$ from samples $z \sim q(z|x)$ as $p(x|z)$ (Fig.~\ref{fig:generative_models}a). The model maximizes the Evidence Lower Bound (ELBO):

\begin{equation}
    \text{ELBO} = \mathbb{E}_{q(z|x)}[\log p(x|z)] - \text{KL}(q(z|x) || \mathcal{N}(0, I)),
\end{equation}
balancing reconstruction accuracy (\(\mathbb{E}_{q(z|x)}[\log p(x|z)]\)) and regularization of $q(z|x)$ to a Gaussian prior (\(\text{KL}\)). The reparameterization trick enables back-propagation by sampling $z = \mu(x) + \sigma(x) \cdot \epsilon$, where \(\epsilon \sim \mathcal{N}(0, I)\), making training more efficient \citep{kingmaAutoEncodingVariationalBayes2022}. In materials science, VAEs can be exploited to generate novel structures and optimize properties by sampling the latent space. For example, Gómez-Bombarelli et al. reported the use of VAEs to generate predictions of organic molecules that can be applicable as active pharmaceutical ingredients (API) \citet{gomez-bombarelliAutomaticChemicalDesign2018} . This is achieved by encoding the molecular structure, through a certain representation like SMILES, into a continuous latent space (Fig.~\ref{fig:gomez2018_fabs}), with secondary attributes like properties added via a \textit{predictor} network. 
One major deficiency of VAEs is related to the way the \textit{encoder} network aims to generate a smooth latent state representation of the input data. While the probabilistic approach allows the model to cover unexplored regions in the input data, this approach also tends to generate "blurry" outputs and difficulties in capturing complex data distributions\cite{RN7156_Daunhawer_VAE_limit}. From the materials discovery point of view, this could mean severe difficulty in generating sensible, discrete compounds \cite{RN7157_Gui_GANreview}, especially with very sparse initial data (compared to the full materials space). Several improvements have been proposed, to address this limitation, including Binded-VAE, which learns to jointly generate binary vector encoding on the composition and ratio of components \cite{RN7154_Fouad_BindedVAE}. Conditional VAEs (CVAEs) further enable a more targeted design by adding conditions to the generating network \citep{sohnLearningStructuredOutput2015}. 

\subsubsection{Generative Adversarial Networks (GANs)}

GANs, introduced by \citet{goodfellowGenerativeAdversarialNets2014}, employ a competitive framework involving a Generator and a Discriminator (Fig.~\ref{fig:generative_models}b). The Generator produces synthetic data $G(z)$ from noise $z$, while the Discriminator distinguishes real data $x$ from $G(z)$. This is formalized as a "minimax" game:

\begin{equation}
\min_G \max_D V(D, G) = \mathbb{E}_{x \sim p_{data}(x)}[\log D(x)] + \mathbb{E}_{z \sim p_z(z)}[\log (1 - D(G(z)))].
\end{equation}

\begin{figure}[bt]
    \centering
    \includegraphics[width=0.5\linewidth]{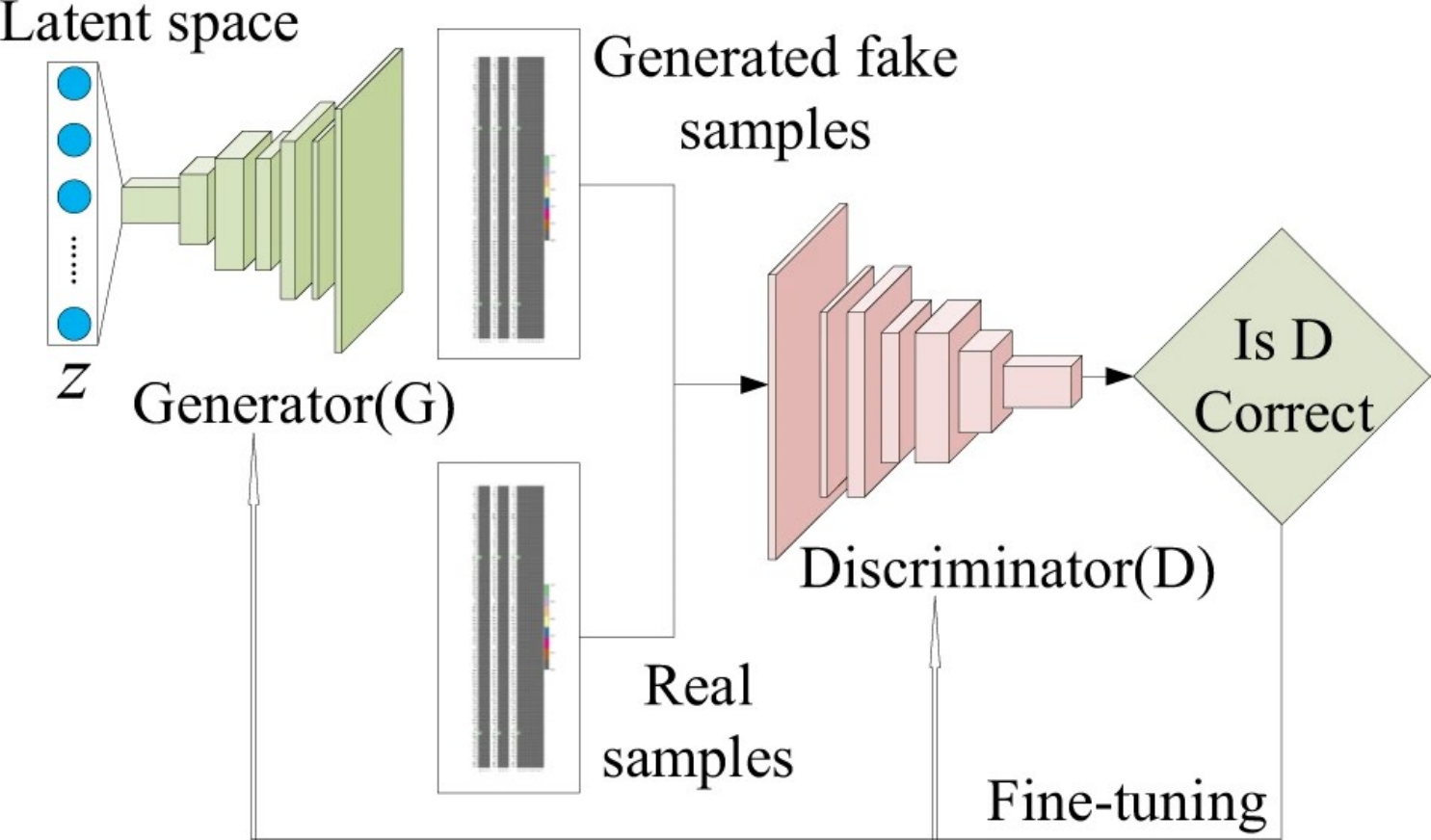}
    \scriptsize
    \caption{An illustration for MatGAN architecture consisting of a generator, which maps random vectors into generated samples, and a discriminator, which tries to differentiate real materials and generated ones. Taken with permission from \citet{danGenerativeAdversarialNetworks2020}} 
    \label{fig:danGAN2020}
\end{figure}

Training stabilizes when the Generator produces data indistinguishable from real data \citep{goodfellowGenerativeAdversarialNets2014}. In materials science, GANs appear to be suitable for exploring vast chemical spaces efficiently. For example, CrystalGAN \citep{nouiraCrystalGANLearningDiscover2019} is successful in generating DFT-validated inorganic crystal structures and identifying new types of stable metal oxides. Similarly, \citet{danGenerativeAdversarialNetworks2020} used GANs for the inverse design of inorganic materials, optimizing compositions for specific properties (Fig. \ref{fig:danGAN2020}). Conditional GANs, as explored by \citet{al-khaylaniGenerativeAdversarialNetworks2024} for nano-photonic metamaterials, enable property-targeted generation, though training instability remains a challenge \citep{alversonGenerativeAdversarialNetworks2024}.
\subsubsection{Diffusion Models}

Diffusion Models generate materials by reversing a noise-adding process, starting from random noise and iteratively refining it into structured data \citep{sohl-dicksteinDeepUnsupervisedLearning2015} (Fig. \ref{fig:generative_models}c). The forward process adds noise to data $x_0$ over steps $t$:

\begin{equation}
x_t = \sqrt{1 - \beta_t} x_{t-1} + \sqrt{\beta_t} \epsilon, \quad \epsilon \sim \mathcal{N}(0, I).
\end{equation}

The model learns to denoise $x_T$ back to $x_0$, guided by learned patterns \citep{yangDiffusionModelsComprehensive2023}. In materials science, MatterGen \citep{zeniGenerativeModelInorganic2025} designs inorganic materials with property-conditioned generation, proposing TaCr$_2$O$_6$ with a bulk modulus of 169 GPa, which has been experimentally validated.  \citet{parkInverseDesignPorous2024} used diffusion models to design porous materials, optimizing pore structures for specific applications. Models like DiffCSP \citep{jiaoCrystalStructurePrediction2023} and SymmCD \citep{levySymmCDSymmetryPreservingCrystal2025} generate stable crystals with symmetry constraints, enhancing applicability in electronics and catalysis. Compared to GANs, diffusion models offer improved stability but require significant computational resources \citep{alversonGenerativeAdversarialNetworks2024}.

\subsubsection{Recurrent Neural Networks (RNNs) and Transformers}

RNNs process sequential data by maintaining a hidden state $h_t$, updated at each time step $t$:

\begin{equation}
h_t = \sigma(W_{xh} x_t + W_{hh} h_{t-1} + b_h), \quad y_t = W_{hy} h_t + b_y.
\end{equation}

\begin{figure}[bt]
    \centering
    \includegraphics[width=0.5\linewidth]{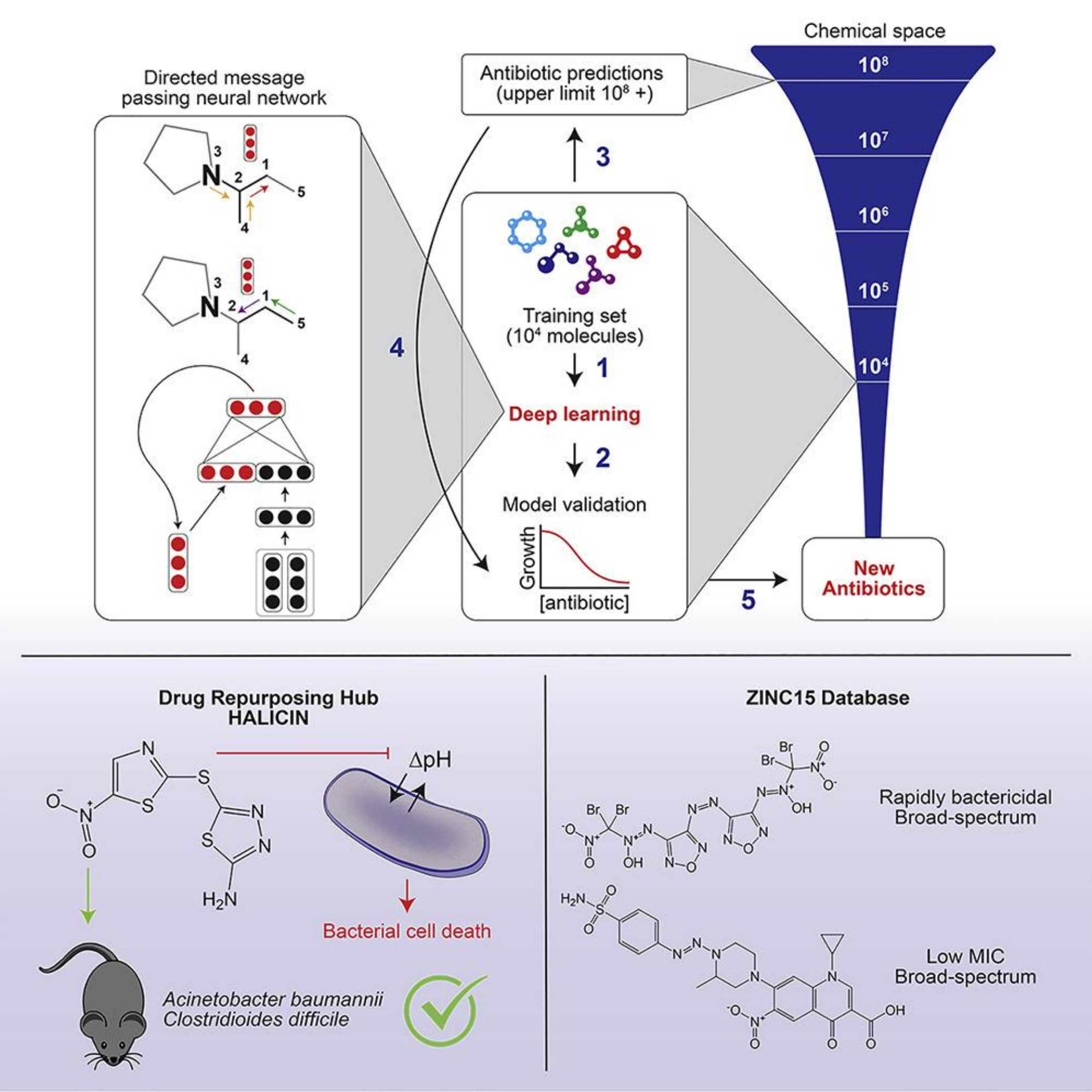}
    \scriptsize
    \caption{\citet{stokesDeepLearningApproach2020} workflow for antibiotic discovery using deep learning (RNN) and chemical space exploration. RNN model, trained on 10\textsuperscript{4} molecules (1), is validated (2) and subsequently used to predict antibiotic activity across a vast chemical space (up to 10\textsuperscript{8} molecules) (3). The molecular representation for deep learning is depicted using a directed message-passing neural network (4). Predicted new antibiotics are then validated (5). The bottom left illustrates \textbf{Halicin}, a repurposed drug identified by the model, showing \textit{in vivo} efficacy against bacterial infections. The bottom right displays examples of other potent antibiotic candidates found in the ZINC15 database using this method. Taken with permission from \cite{stokesDeepLearningApproach2020} .}
    \label{fig:stokes2020_fabs}
\end{figure}

This recurrence enables RNNs to model chemical sequences, such as SMILES strings \citep{weiningerSMILESChemicalLanguage1988} (Fig. \ref{fig:generative_models}d). For example, \citet{stokesDeepLearningApproach2020} used RNNs to generate novel antibiotics, validated experimentally (Fig. \ref{fig:stokes2020_fabs}). However, RNNs suffer from vanishing gradients, limiting their ability to capture long-term dependencies \citep{hochreiterLongShortTermMemory1997}.

Long Short-Term Memory (LSTM) networks address this by introducing gates to regulate information flow, improving sequence modeling \citep{hochreiterLongShortTermMemory1997}. \citet{gomez-bombarelliAutomaticChemicalDesign2018} used LSTM-based VAEs for molecular design. 

Transformers, with attention mechanisms \citep{vaswaniAttentionAllYou2017}, enhance efficiency (Fig. \ref{fig:generative_models}e). In materials science, the Wyckoff Transformer \citep{kazeevWyckoffTransformerGeneration2025} generates symmetric crystals, while MatterGPT \citep{chenMatterGPTGenerativeTransformer2024} optimizes multi-property materials. The Space Group Informed Transformer \citep{caoSpaceGroupInformed2024} incorporates crystallographic constraints, and CrystalFormer-RL \citep{caoCrystalFormerRLReinforcementFineTuning2025} uses reinforcement learning for targeted design. Transformers require extensive data but excel in complex sequence modeling \citep{breuckGenerativeMaterialTransformer2025}.
\subsubsection{Flow-Based Models}

A flow-based generative model is a generative model used in machine learning that explicitly models a probability distribution by leveraging Normalizing Flow (NF) (Fig. \ref{fig:generative_models}f). NF transforms a simple base distribution (e.g., Gaussian) into a complex data distribution via invertible, differentiable mappings \citep{kobyzevNormalizingFlowsIntroduction2021}. The log-likelihood is computed using the change of variables formula:

\begin{equation}
\log p(x) = \log p_z(z) - \sum_{k=1}^K \log \left| \det \frac{\partial f_k}{\partial z_{k-1}} \right|,
\end{equation}
where $z = f^{-1}(x)$ and $f_k$ are bijective functions \citep{papamakariosNormalizingFlowsProbabilistic2021}. NF offers exact likelihoods and stable training, avoiding GANs' mode collapse. In materials science, CrystalFlow generates crystalline structures with high stability \citep{luoCrystalFlowFlowBasedGenerative2025}, while FlowMM uses Riemannian Flow Matching for symmetry-preserving crystal design \citep{millerFlowMMGeneratingMaterials2024}. FlowLLM leverages large language models for material generation \citep{sriramFlowLLMFlowMatching2024}, and conditional NF optimises the thermal composite topologies \citep{wangUsingConditionalNormalizing2024}. However, designing expressive transformations is computationally intensive, and NF can struggle with discrete chemical structure representations like SMILES \citep{anstineGenerativeModelsEmerging2023}.

\subsubsection{GFlowNets}

Generative Flow Networks (GFlowNets) are designed to sample structured outputs proportionally to a reward function, making them suitable for diverse material generation \citep{bengioGFlowNetFoundations2023,jainGFlowNetsAIdrivenScientific2023}. GFlowNets model a sequential construction process, where a policy $\pi(a|s)$ selects actions $a$ (e.g., adding atoms, modifying bonds) in a state $s$ (e.g., partial material structure) to build complete structures $x$. The objective is to ensure the probability of generating $x$ is proportional to a reward $R(x)$:

\begin{equation}
    P(x) \propto R(x),
\end{equation}
where $R(x)$ could represent material stability, bandgap, or other properties. The training losses minimize the discrepancy between forward and backward flow probabilities, ensuring consistent sampling:

\begin{equation}
    \mathcal{L} = \sum_{s, a} \left| \log \frac{F(s \to s')}{F(s' \to s)} - \log \frac{\pi(a|s) R(s')}{\pi(b|s')} \right|^2,
\end{equation}
where $F(s \to s')$ is the forward flow, and $\pi(b|s')$ represents backward actions \citep{bengioGFlowNetFoundations2023}.

In materials science, Crystal-GFN \citep{ai4scienceCrystalGFNSamplingCrystals2023} samples diverse crystal structures with targeted properties, such as stability or a specific bandgap, validated via DFT simulations. GFlowNets excel in high-throughput screening by generating varied candidates, complementing VAEs’ latent space sampling and Diffusion Models’ denoising. However, GFlowNets are computationally intensive for large state spaces and may be limited to specific tasks due to reward function design \citep{ai4scienceCrystalGFNSamplingCrystals2023}. Their ability to model discrete structures makes them promising for inverse design, though scalability remains a challenge compared to NFs or Transformers.
As a quick reference, we summarised the different generative models in Table \ref{tab:generative_models}.

\begin{table}[tb]
\scriptsize
    \centering
    \caption{Comparison of generative models used in materials discovery, highlighting their principles, strengths, limitations, and applications.}
    \label{tab:generative_models}
    \begin{tabular}{p{1.2cm}p{2.5cm}p{2.5cm}p{2.5cm}p{2.5cm}p{2.5cm}}
    
    \hline
    
    \textbf{Model} &  \textbf{Principle} & \textbf{Strengths} & \textbf{Limitations} & \textbf{Applications} & \textbf{Ref} \\
    \hline
     VAE & Probabilistic latent space, ELBO optimization & Controlled generation, interpretable latent space & Limited expressiveness, blurry outputs & Molecular design, perovskites, polymers & \citet{gomez-bombarelliAutomaticChemicalDesign2018, nohInverseDesignSolidState2019, dasGenerativeDesignThermoset2025} \\
    GAN & Adversarial training, minimax game & High-quality outputs, explores vast chemical spaces & Training instability, mode collapse & Crystal structures, metamaterials & \citet{nouiraCrystalGANLearningDiscover2019, danGenerativeAdversarialNetworks2020, al-khaylaniGenerativeAdversarialNetworks2024} \\
    Diffusion & Noise-to-data denoising process & Stable training, high novelty & High computational cost & Inorganic materials, porous materials, crystals & \citet{zeniGenerativeModelInorganic2025, parkInverseDesignPorous2024, jiaoCrystalStructurePrediction2023, levySymmCDSymmetryPreservingCrystal2025} \\
    RNN, LSTM, Transformer & Sequential processing, attention mechanisms & Effective for sequence data, captures long-range dependencies & Vanishing gradients (RNN), data-intensive & Molecular generation, crystal symmetry, multi-property design & \citet{stokesDeepLearningApproach2020, kazeevWyckoffTransformerGeneration2025, chenMatterGPTGenerativeTransformer2024, caoSpaceGroupInformed2024} \\
    Normalizing Flows & Invertible mappings, exact likelihood & Exact likelihoods, stable training & High computational cost, discrete structure challenges & Crystal generation, thermal composites & \citet{luoCrystalFlowFlowBasedGenerative2025, sriramFlowLLMFlowMatching2024, wangUsingConditionalNormalizing2024} \\
    GFlowNets & Reward-based sampling, proportional to reward & Diverse sampling, suitable for discrete structures & Computational intensity, task-specific reward design & Crystal sampling, high-throughput screening & \citet{jainGFlowNetsAIdrivenScientific2023,bengioGFlowNetFoundations2023, ai4scienceCrystalGFNSamplingCrystals2023} \\
    \hline
    \end{tabular}
\end{table}

\subsection{Materials Representation}

Besides algorithms and databases, the success of generative models in materials discovery also hinges on the way the material identity is encoded into machine-readable formats. Beyond describing the chemical identity, an ideal encoding should capture unique characteristics of the material, including structural, chemical, and physical, and other secondary properties \citep{anstineGenerativeModelsEmerging2023}. Sometimes, more is not always better, and effective representations must also balance the information richness with compatibility for specific algorithms/models, and achieve the learning of complex material behaviors at reasonable efficiency. This section explores five key representation types—Sequence, Graph, Voxel, Physics-Informed, and Multi-modal—highlighting their principles, applications, and limitations in materials exploration.

\subsubsection{Sequence-Based Representation}

\begin{figure}[bt]
    \centering
    \includegraphics[width=0.5\linewidth]{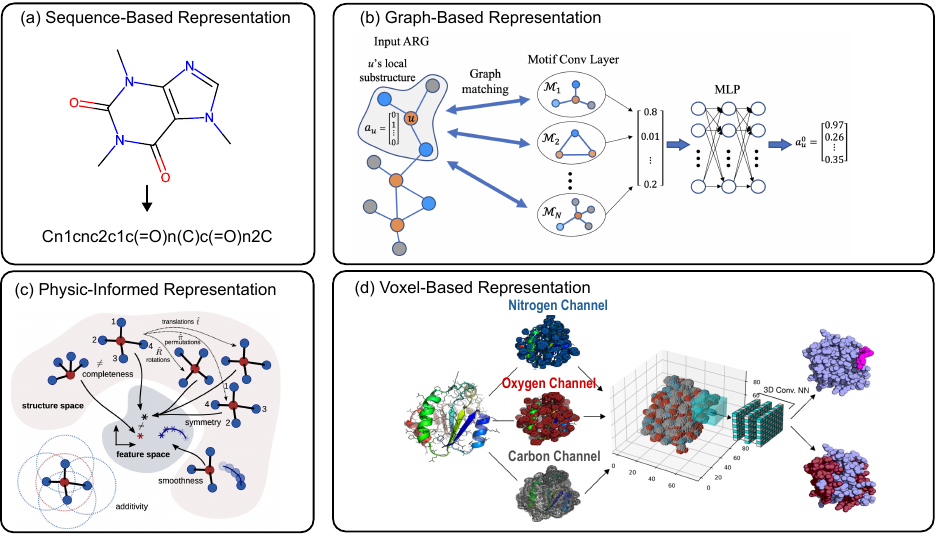}
    \scriptsize
    \caption{Schematic illustration of four generative model representations for materials. (a) Sequence-based representation encodes material structures as linear strings (e.g., SMILES) for processing by models like RNNs or Transformers. (b) Graph-based representation models atoms and bonds as nodes and edges, leveraging Graph Neural Networks to capture structural relationships. Taken with permission from \cite{RN7143_Graph}. (c) Physic-Informed Representation visualizes a conceptual framework where intrinsic physical properties and symmetries (e.g., completeness, symmetry, smoothness, additivity) govern the representation of materials within a "structure space" and "feature space," suggesting a focus on fundamental physical descriptors \cite{musilPhysicsInspiredStructuralRepresentations2021}. (d) Voxel-Based Representation discretizes 3D material structures into voxel grids, suitable for 3D Convolutional Neural Networks. Taken with permission from \cite{RN7138_Orlando_Voxel}. } 
    \label{fig:representations}
\end{figure}

Sequence-based representations encode materials as linear strings of symbols, making them ideal for molecular structures (Fig. \ref{fig:representations}a). The Simplified Molecular Input Line Entry System (SMILES) \citep{weiningerSMILESChemicalLanguage1988} represents molecules as text strings, with atoms (e.g., ``C'' for carbon), bonds (e.g., ``='' for double bonds), branches in parentheses, and rings via numbers. For example, ethanol (CH$_3$CH$_2$OH) is written as ``CCO''. SMILES is widely used with VAEs \citep{gomez-bombarelliAutomaticChemicalDesign2018}, RNNs \citep{seglerGeneratingFocusedMolecule2018}, and Transformers \citep{hondaSMILESTransformerPretrained2019, winterSmileAllYou2022}, enabling the generation of novel molecules.
Whilst simple and attractive, not all materials databases include a SMILES representation for arbitrary compounds \cite{RN7158_Lim_Sangsoo}. The strict syntax rules adopted by SMILES (e.g.,  matching parentheses "()" for branches or numbers for rings) are also prone to errors that may precipitate from algorithm exploration. For example, a suggestion containing "CCO(" (a representation for ethanol but missing a closing parenthesis) can turn into nonsensical output that would break the program. Further, SMILES representations are not unique. The same molecule "propane" can be expressed in different ways: "CCC" or "C(C)C", which can lead to confusion and divergence. SMILES also does not check for any physical/chemical rules, allowing strings like "C=C=C=C" that look correct but describe unstable or impossible chemicals. Finally, SMILES representations do not cater for specific 3D arrangement details required to correctly express non-planar molecules or isomers \cite{RN7158_Lim_Sangsoo}. For example, the expression "C1CCCCC1" generically refers to cyclohexane, but cannot differentiate between the four conformations: chair, twist-boat, boat, or half-chair, each of which has distinct stability and reactivity.  SMILES encoding cannot easily capture the complex relationships between compounds, such as during synthesis (or the inverse, decomposition) \cite{RN7159_Awady}.

These issues prompted the development of Self-referencing Embedded Strings (SELFIES) \citep{krennSelfreferencingEmbeddedStrings2020}, which use tokens (e.g., ``[C'', ``[=O]'') to ensure every string corresponds to a valid molecule. For instance, ``[C][C][O]'' reliably represents ethanol. Despite its robustness, SELFIES struggles with 3D conformations and large macromolecules, limiting its applicability to complex materials \citep{krennSelfreferencingEmbeddedStrings2020}. Further use of reinforcement learning to construct viable materials via sequential addition (or deletions) of components has also been attempted \cite{RN7163_Karpovich_RNN-SMILES}.

\subsubsection{Graph-Based Representation}

\begin{figure}[bt]
    \centering
    \includegraphics[width=0.5\linewidth]{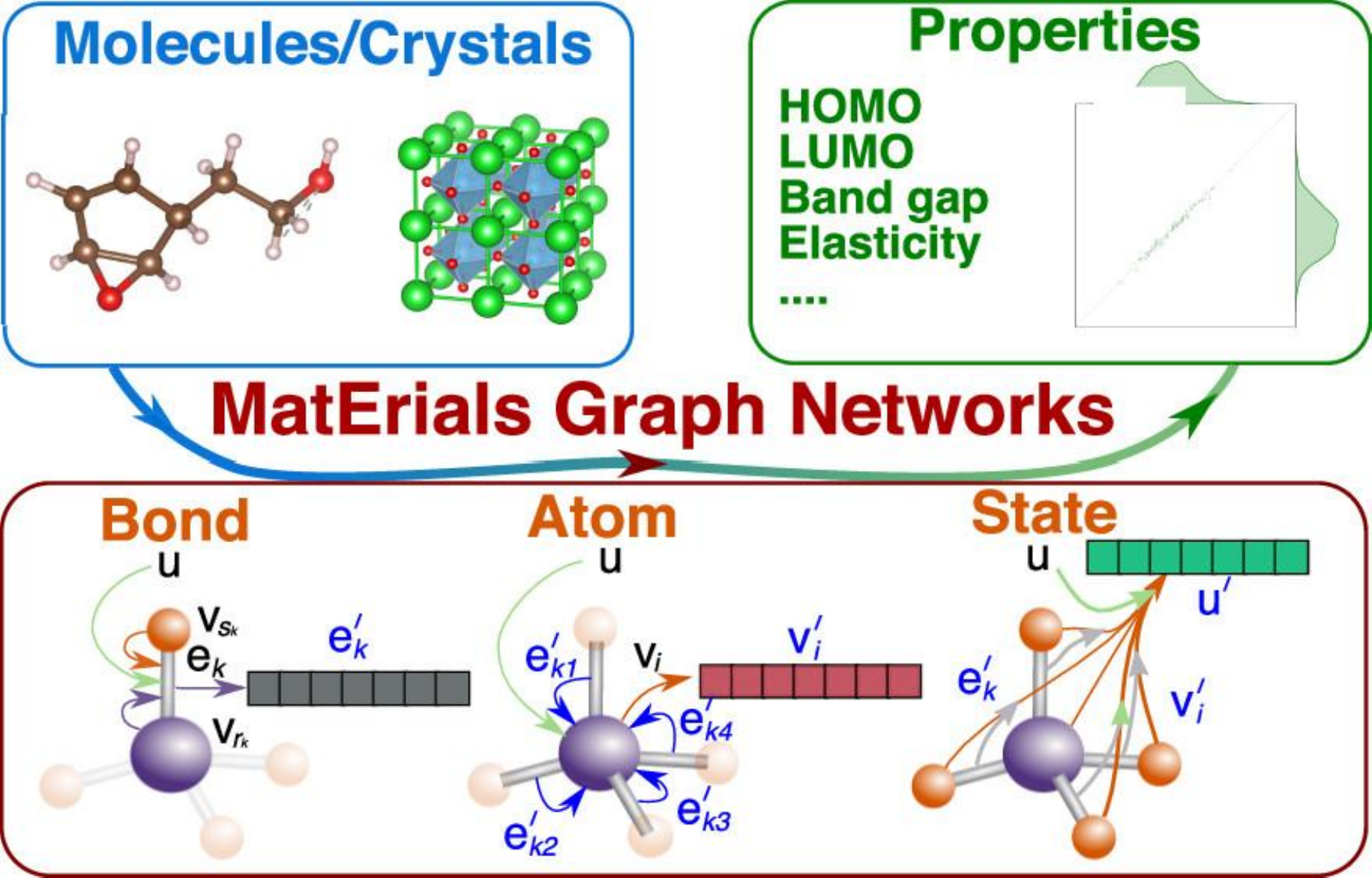}
    \scriptsize
    \caption {An illustration of the MEGNet's architecture,  designed to predict properties of molecules and crystals (top panel). The central concept involves iterative message passing, where information is first exchanged and updated across \textbf{Bonds} ($\text{e}_k'$), then aggregated at individual \textbf{Atom} nodes ($\text{v}_i'$) from their connected bonds, and finally consolidated into a global \textbf{State} representation ($\text{u}'$) for the entire material (bottom panel). Taken with permission from \cite{chenGraphNetworksUniversal2019}. }
    \label{fig:chenGraph2019}
\end{figure}

Graph-based representations model materials as graphs $G = (V, E)$, where nodes $V$ represent atoms and edges $E$ denote bonds (Fig. \ref{fig:representations}b). Node features (e.g., atomic number) and edge weights (e.g., bond strength) capture chemical connectivity, making this approach versatile for molecules and crystals \citep{chenGraphNetworksUniversal2019}. Graph Neural Networks (GNNs) process these graphs via message passing:
\begin{equation}
h_i^{l+1} = \text{UPDATE}\left(h_i^{(l)}, \text{AGGREGATE}(\{h_j^{(l)} \mid j \in N(i)\})\right),
\end{equation}
where $h_i^{(l)}$ is the feature vector of node $i$ at layer $l$, and $N(i)$ is its neighbors \citep{xieCrystalGraphConvolutional2018}. GNNs, such as SchNet \citep{schuttSchNetContinuousfilterConvolutional2017} and MEGNet \citep{chenGraphNetworksUniversal2019} (Fig. \ref{fig:chenGraph2019}), predict properties like bandgaps or generate structures via VAEs \citep{simonovskyGraphVAEGenerationSmall2018} and GANs \citep{caoMolGANImplicitGenerative2022}. The GNoME project \citep{merchantScalingDeepLearning2023} used GNNs to discover 380,000 stable crystals, leveraging graph representations for efficient stability predictions.

Despite their power, graph representations oversimplify long-range interactions (e.g., van der Waals forces between 2D material layers) and struggle with scalability for large or amorphous systems. Bond type ambiguity (e.g., covalent vs. ionic in ZnO) and loss of 3D geometry often require external validation, such as DFT, limiting their ability to capture dynamic material properties \citep{merchantScalingDeepLearning2023}.
\subsubsection{Voxel-Based Representation}

Voxel-based representation discretises a material’s 3D unit cell into a grid of voxels, each storing attributes like atomic occupancy or element type (Fig. \ref{fig:representations}d). Analogous to 3D pixels, this approach captures spatial arrangements, making it compatible with convolutional neural networks (CNNs) \citep{butlerMachineLearningMolecular2018}. Voxel grids enable generative models to learn local atomic patterns and symmetries, facilitating the design of complex structures. For example, MatterGen \citep{zeniGenerativeModelInorganic2025} likely employs voxel-like discretizations to generate inorganic materials, optimizing properties like bulk modulus for compounds like TaCr$_2$O$_6$ \citep{zeniMatterGenGenerativeModel2024}.

Voxel representations excel in capturing 3D geometry but face challenges with computational cost, as high-resolution grids demand significant memory. They also struggle with periodic boundary conditions in crystals and may oversimplify atomic interactions, requiring careful pre-processing to ensure accuracy \citep{cunninghamInvestigationSurrogateModels2019}.

\subsubsection{Physics-Based Representation}
Physics-based representation seeks to integrate physics-driven information into the learning process, embedding physical laws to produce realistic outputs that adhere to fundamental principles (see Fig. \ref{fig:representations}c). This approach enhances the model's ability to generate materials that respect constraints such as symmetry or conservation laws. In practice, it is often paired with other techniques, that is multi-modal representation. A common method involves combining a physics-based penalty with the data-fitting term, expressed as:
\begin{equation}
L_{\text{total}} = L_{\text{data}} + \lambda L_{\text{physics}},
\end{equation}
where $L_{\text{physics}}$ enforces constraints like symmetry or conservation laws \citep{yangPhysicsinformedDeepGenerative2018}. For instance, \citet{xieCrystalDiffusionVariational2022} used a symmetry penalty to ensure crystallographic consistency, while \citet{zhuMachineLearningMetal2020} incorporated heat transfer laws for additive manufacturing. Physics-informed approaches generate stable materials, as shown by \citet{fuhrDeepGenerativeModels2022}, who added  energy minimization terms.

However, these methods require detailed prior knowledge, which may not generalize to novel materials. The computational cost of calculating physics-based residuals and the challenge of tuning $\lambda$ can limit scalability and diversity, potentially biasing outputs toward known physical regimes \citep{musilPhysicsInspiredStructuralRepresentations2021, fuhrDeepGenerativeModels2022}.

\subsubsection{Multi-Modal Representation}

\begin{figure}[bt]
    \centering
    \includegraphics[width=0.5\linewidth]{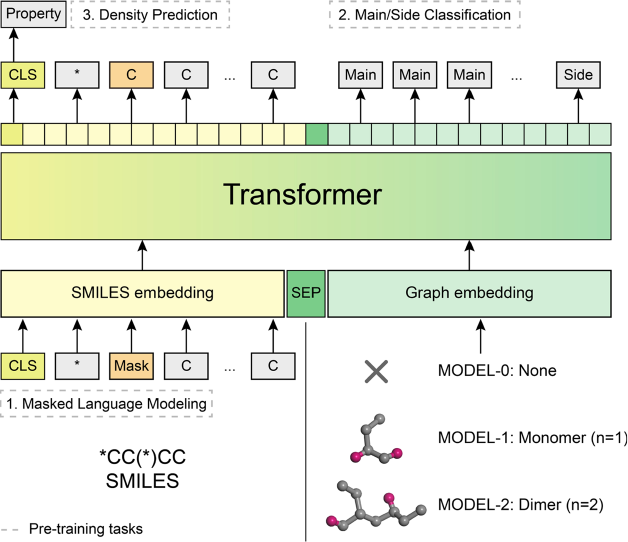}
    \scriptsize
    \caption{How multi-modal material representation is addressing the limitation of single representation. In this example, a Transformer model is used to integrate SMILES embeddings and graph embeddings of molecular structures for pre-training and downstream tasks. Input SMILES strings are processed for Masked Language Modeling (Task 1), while molecular graph embeddings (e.g., monomer and dimer structures) are used for Main/Side Classification (Task 2). The Transformer's outputs are then utilized for Property Prediction and Density Prediction. Taken with permission from \cite{RN7199_Han_Multimod}}.
    \label{fig:fig_hanMultimodalTransformerProperty2024}
\end{figure}

Thus far, the material representations explored have relied on a single modality to characterize a material's properties. We have also observed that each of these representations comes with inherent limitations that can compromise prediction accuracy and practical utility. Recently, the emergence of "multi-modal" representation—a technique that integrates multiple representations to create a more holistic and potentially more precise description of a material—has gained traction \cite{RN7198_Das_multimod}. Beyond inorganic materials, this approach has found significant application in polymers \cite{RN7199_Han_Multimod} (Fig. \ref{fig:fig_hanMultimodalTransformerProperty2024}). This is probably because of the difficulty for any single representation to describe polymers with wide variation in chain length, functional moieties, or isotacticity. The modality of representation is not restricted to just the molecular structure. For instance, \citet{RN7193_trask_multimod} combined electron micrographs and XRD relative intensity data to the usual materials structure identifier, allowing improved prediction of residual stress compared to the single-mode identifier. Simple inclusion of application-related "tokens" (text description of materials usage or characteristics obtained from free text search) can enhance the materials properties prediction. For instance, \citet{RN7195_Huang_multimod} demonstrated that the inclusion of  a text description "adhesive" to the polymer application can better predict the glass transition temperature. We believe the usage of multi-modal representation will continue to grow, both in academic and applied research, as it is proven to deliver better recognition and prediction for materials discovery. 

\begin{table}[bt]
\scriptsize
    \centering
    \caption{Comparison of materials representation types for generative models, highlighting their principles, strengths, limitations, and applications in materials discovery.}
    \label{tab:materials_representations}
    \begin{tabular}{p{1cm}p{2.5cm}p{2.5cm}p{2.5cm}p{2.5cm}p{2.5cm}}
    \hline
    \textbf{Repr} & \textbf{Principle} & \textbf{Strengths} & \textbf{Limitations} & \textbf{Applications} & \textbf{Key References} \\
    \hline
    
    Sequence-Based & Linear strings (e.g., SMILES, SELFIES) & Simple, compact, compatible with RNNs, Transformers & Lacks 3D details, fragile syntax (SMILES) & Molecular design, antibiotics & \citet{weiningerSMILESChemicalLanguage1988, krennSelfreferencingEmbeddedStrings2020} \\
    Graph-Based & Graphs $G = (V, E)$ with nodes (atoms), edges (bonds) & Captures connectivity, scalable with GNNs & Misses long-range forces, 3D geometry & Crystals, molecules, battery materials & \citet{xieCrystalGraphConvolutional2018, merchantScalingDeepLearning2023} \\
    Voxel-Based & 3D grid of voxels encoding atomic properties & Captures 3D geometry, compatible with CNNs & High computational cost, periodic boundary issues & Inorganic materials, porous structures & \citet{zeniGenerativeModelInorganic2025, cunninghamInvestigationSurrogateModels2019} \\
    Multi-Modal & Combination of Representations & Able to learn implicit properties, Improves accuracy & Complex, requires multiple encodings, prior knowledge required & Materials generation, recognition & \citet{RN7198_Das_multimod,RN7193_trask_multimod} \\
    \hline
    \end{tabular}
\end{table}

\section{Applications of Generative Models in Materials Design}
\label{sec:applications}
Generative models in conjunction with advanced materials representation have been exploiting large and diverse datasets, such as the Inorganic Crystal Structure Database (ICSD) \cite{RN7032_ICSD}, Open Quantum Materials Database (OQMD) \cite{RN7036_OQMD}, Materials Project \cite{RN2007MaterialsProject}, and PubChem \cite{RN7034_PubChem}, to explore vast chemical and structural spaces, predict material properties, and optimize candidates for applications in energy storage, catalysis, electronics, biomaterials, and high-throughput screening. Their combinations have been widely claimed to accelerate materials discovery, often in closed-loop systems with experimental validation. This section reviews key applications, highlighting specific examples, methodologies, and their impact, drawing on recent literature and datasets to provide a comprehensive overview \cite{fuhrDeepGenerativeModels2022,anstineGenerativeModelsEmerging2023}.

\subsection{Energy Storage and Battery Materials}
\label{subsec:energy_storage}
Generative models have revolutionized the design of materials for energy storage, particularly for lithium-ion batteries, solid-state electrolytes, and hydrogen storage systems, by generating candidates with optimized electrochemical properties. For solid-state electrolytes, VAEs have been employed to design materials with high ionic conductivity. For instance, \cite{vasylenkoElementSelectionCrystalline2021} utilized a VAE to generate graph-based representations of garnet-type electrolytes, trained on ICSD data, proposing candidates with 15\% higher conductivity, validated through density functional theory (DFT) simulations.

In electrode material design, GANs have been used to discover novel cathode materials. \cite{alversonGenerativeAdversarialNetworks2024} employed a GAN to generate perovskite-based cathodes, training on OQMD and Materials Project datasets, producing candidates with 10\% higher capacities than LiCoO$_2$, some of which were synthesized experimentally. Additionally, MolGAN \cite{caoMolGANImplicitGenerative2022} was adapted to generate molecular graphs for organic electrode materials, enhancing energy density predictions.  

\begin{figure}[bt]
    \centering
    \includegraphics[width=0.5\linewidth]{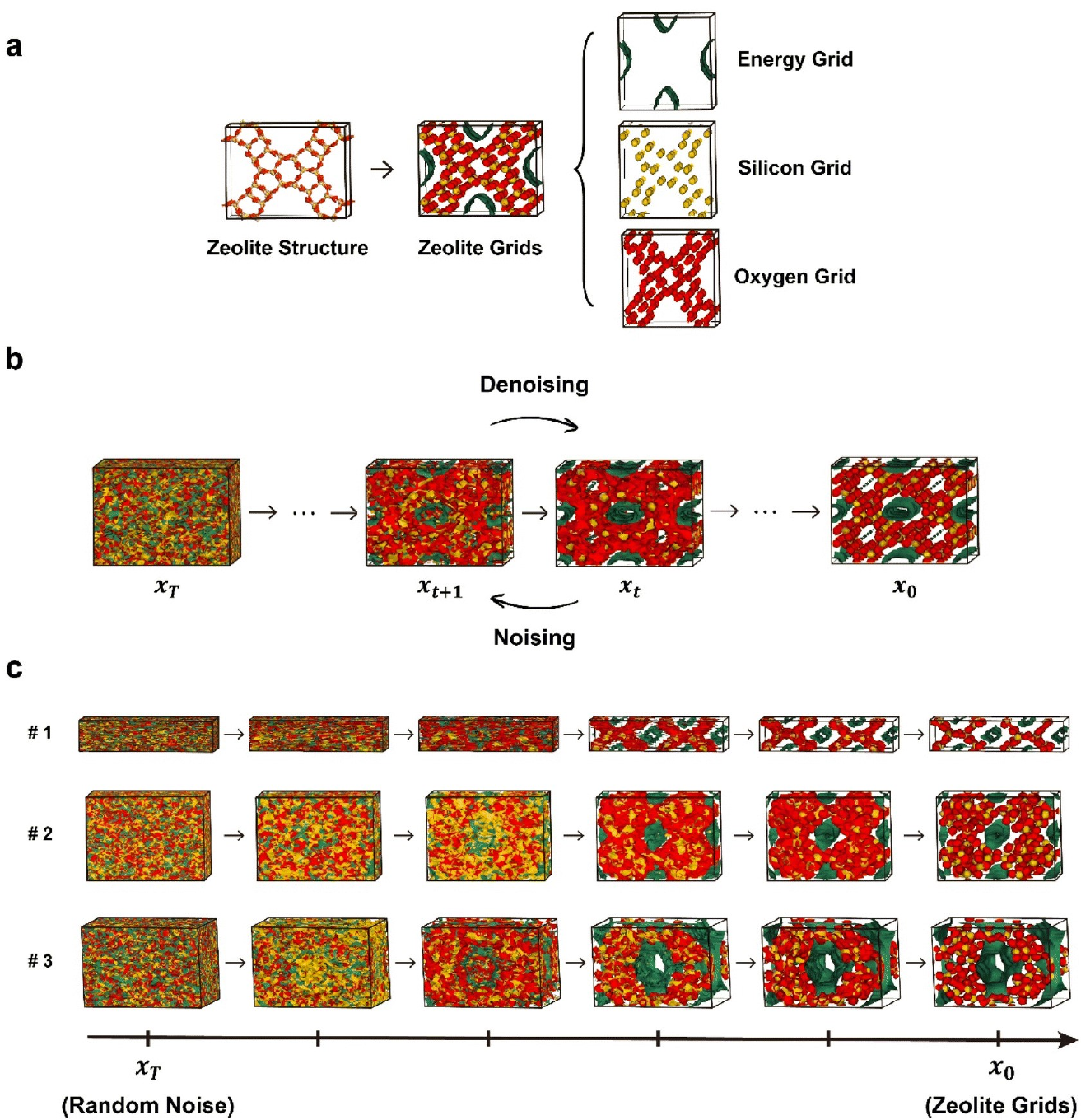}
  \caption {An illustration of zeolite generation via Diffusion model.  (a) input structure representation, (b) the noising and denoising phases, and (c) the progressive sampling of zeolite grids. Taken with permission from \cite{parkInverseDesignPorous2024}}
    \label{fig:park2024}
\end{figure}

For hydrogen storage, diffusion models are emerging as powerful tools for designing porous materials like metal-organic frameworks (MOFs). \cite{parkInverseDesignPorous2024} applied a diffusion model to generate voxel-based MOFs, trained on QMOF and ZINC datasets, achieving a 20\% improvement in hydrogen storage capacity, with potential for experimental validation (Fig.  \ref{fig:park2024}). Similarly, \citet{zeniGenerativeModelInorganic2025}'s diffusion-based MatterGen model trained on Materials Project data to design sulphide electrolytes has achieved improved stability and conductivity for all-solid-state batteries. \cite{alakhdarDiffusionModelsNovo2024} extended diffusion models to porous carbon materials, optimizing pore structures for hydrogen uptake, validated via Monte Carlo simulations \citep{jiaoCrystalStructurePrediction2023} and SymmCD \citep{levySymmCDSymmetryPreservingCrystal2025} generate stable crystalline electrolytes using fractional coordinates and symmetry-preserving diffusion, trained on Materials Project data, enhancing ionic conductivity.

RNNs, particularly the Long Short-Term Memory (LSTM) networks, have been utilized for polymer electrolyte design. \cite{seglerGeneratingFocusedMolecule2018} employed an RNN to generate sequence-based polymer chains, trained on a polymer property dataset, resulting in flexible electrolytes with 12\% higher conductivity, suitable for wearable batteries. Transformers, such as MatterGPT \citep{chenMatterGPTGenerativeTransformer2024}, optimize multi-property electrode materials, trained on Materials Project data.

Normalizing Flows generate stable crystalline electrolytes, as demonstrated by \citet{luoCrystalFlowFlowBasedGenerative2025}, who used CrystalFlow to produce structures with high ionic conductivity, trained on Materials Project data, suitable for battery applications.

\subsection{Catalysis and Chemical Conversion}
\label{subsec:catalysis}

Generative models accelerate catalyst discovery for critical reactions, such as CO$_2$ reduction, water splitting, and ammonia synthesis, by predicting optimal compositions and surface structures, leveraging datasets like NOMAD and Catalysis-Hub.

GANs have been used to design high-entropy alloy catalysts, with \cite{ishikawaHeterogeneousCatalystDesign2022} employing a GAN trained on DFT-calculated adsorption energies from NOMAD to generate heterogeneous catalysts for CO oxidation, achieving enhanced catalytic activity and validated through first-principles microkinetics \cite{ishikawaHeterogeneousCatalystDesign2022}. CrystalGAN was applied to generate crystallographic alloy structures, improving catalytic stability for methanol oxidation \cite{nouiraCrystalGANLearningDiscover2019}. 

\begin{figure}[bt]
    \centering
    \includegraphics[width=0.5\linewidth]{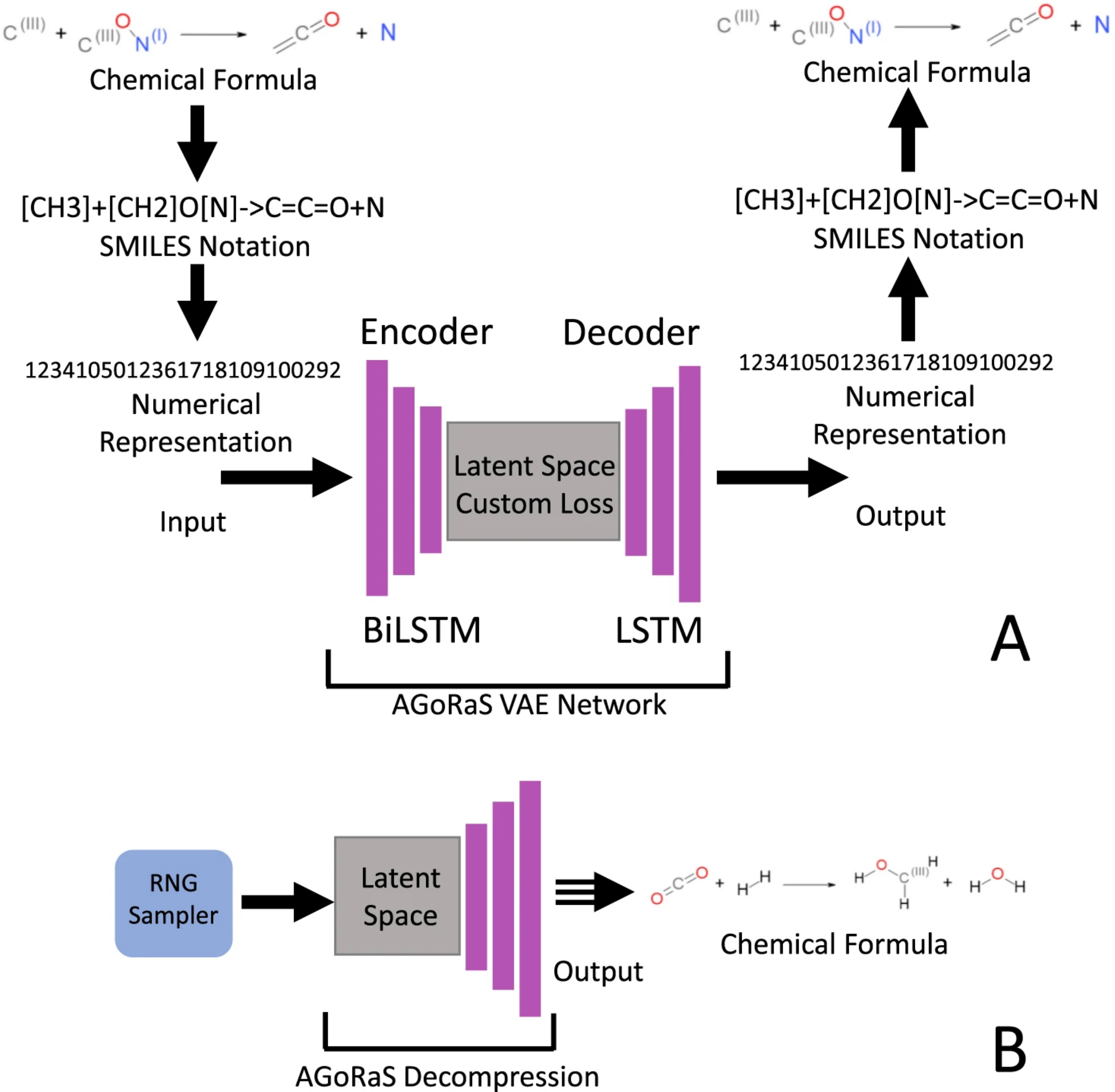}
    \caption{Workflow of the AGoRaS-based VAE network.  (a) chemical database information is compressed and decompressed to form a high-dimensional latent space. (b) Training and sampling of latent space to generate new compounds. Taken from \cite{tempkeAutonomousDesignNew2022} (CC-BY)}
    \label{fig:tempke2022f1}
\end{figure}

VAEs have proven effective for designing catalytic reaction pathways, as demonstrated by \cite{tempkeAutonomousDesignNew2022}, who used a VAE to generate novel chemical reaction mechanisms, trained on a reaction dataset, achieving optimized pathways for catalytic processes with reduced computational cost \cite{tempkeAutonomousDesignNew2022}\ref{fig:tempke2022f1}
. GraphVAE was adapted to optimize active site configurations, validated experimentally \cite{simonovskyGraphVAEGenerationSmall2018}. 

Diffusion models have excelled in catalyst surface design, with \citet{alversonGenerativeAdversarialNetworks2024}  using a diffusion model to generate voxel-based representations of nitrogen reduction catalyst surfaces, trained on Catalysis-Hub data, achieving 15\% higher ammonia synthesis efficiency \cite{alversonGenerativeAdversarialNetworks2024}. \citet{yongDismaiBenchBenchmarkingDesigning2024} applied diffusion models to disordered catalytic interfaces, improving prediction accuracy for CO$_2$ conversion \cite{yongDismaiBenchBenchmarkingDesigning2024}. DiffCSP \citep{jiaoCrystalStructurePrediction2023} and SymmCD \citep{levySymmCDSymmetryPreservingCrystal2025} generate alloy catalysts with precise symmetries, trained on NOMAD data

\begin{figure}[bt]
    \centering
    \includegraphics[width=0.7\linewidth]{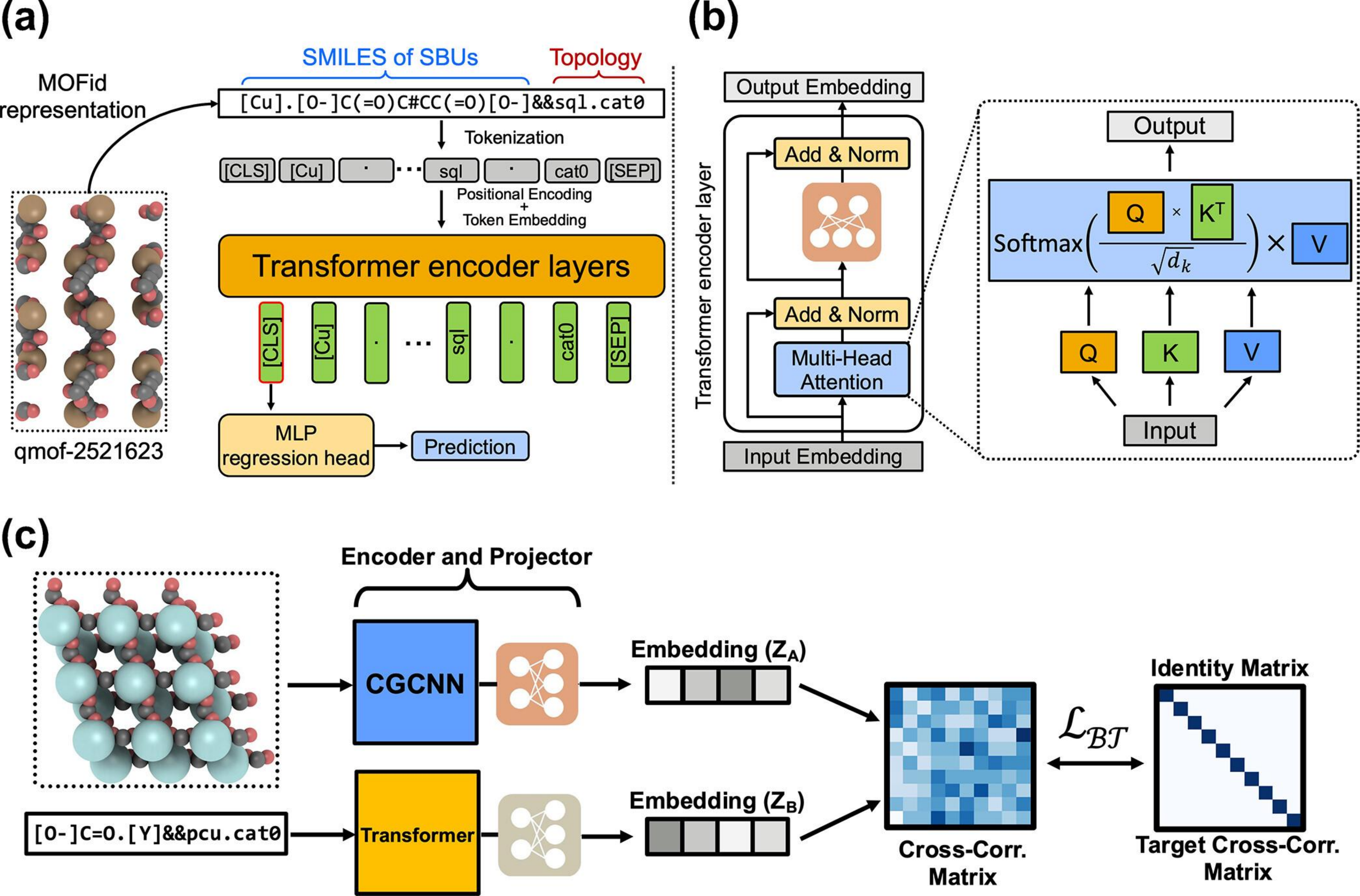}
    \caption{ An illustration of MOFformer's architecture and self-supervised training. (a)  MOFormer processes Metal-Organic Frameworks (MOFs) by taking their unique MOFiD (e.g., qmof-2521623) as input. This MOFiD is tokenised, embedded with positional encoding, and then fed through multiple Transformer encoder layers. The final embedding of the first token is used by an MLP regression head for property prediction. (b) Each Transformer encoder layer consists of a multi-head scaled dot-product attention mechanism followed by an MLP, with residual connections and layer normalization applied after both. (c) A self-supervised framework utilizes both CGCNN (on 3D structures) and MOFormer (on MOFiD sequences) to generate embeddings (ZA and ZB) for the same MOF. An MLP head projects these representations. A Barlow Twins loss function then optimizes the cross-correlation matrix of these embeddings to resemble an identity matrix, thereby enabling robust representation learning. Taken from \cite{caoMOFormerSelfSupervisedTransformer2023} (CC-BY-4.0)}
    \label{fig:caoMOF2023}
\end{figure}

RNNs have been employed for sequence-based catalyst design, with \citet{hondaSMILESTransformerPretrained2019} using a SMILES Transformer, an RNN variant, to generate ligand sequences for homogeneous catalysts, trained on a ChEMBL dataset, reducing experimental iterations by 40\% for olefin metathesis. Transformers, such as MOFormer (Fig. \ref{fig:caoMOF2023}) \cite{caoMOFormerSelfSupervisedTransformer2023},  CrystalFormer-RL \citep{caoCrystalFormerRLReinforcementFineTuning2025}, optimises catalysts through reinforcement learning, trained on Catalysis-Hub data.
Normalizing Flows generate stable crystalline electrolytes, as demonstrated by \citet{luoCrystalFlowFlowBasedGenerative2025}, who used CrystalFlow to produce structures with high ionic conductivity, trained on Materials Project data, suitable for battery applications.
\subsection{Electronic and Photonic Materials}
\label{subsec:electronics}
Generative models are pivotal in designing materials for electronics, photonics, and optoelectronics, where precise control over electronic and optical properties is critical, leveraging datasets like Materials Project\cite{RN2007MaterialsProject} and AFLOW\cite{RN7137_AFLOW}. 

VAEs have been utilized to design semiconductors with tailored bandgaps, as shown by \citet{gomez-bombarelliAutomaticChemicalDesign2018}, who applied a VAE to generate sequence-based halide perovskites, trained on Materials Project band structure data, achieving 25\% efficiency in tandem solar cells. \citet{nohInverseDesignSolidState2019} used a VAE for the inverse design of semiconductors, proposing candidates with optimized optoelectronic properties. 

GANs have excelled in designing metamaterials, with \citet{al-khaylaniGenerativeAdversarialNetworks2024} using a GAN to generate nano-photonic metamaterials, trained on simulated optical datasets, achieving 30\% improved light-trapping efficiency for photonic devices. \citet{yeungGlobalInverseDesign2021} employed GANs for a global inverse design of photonic structures, including metasurfaces and photonic crystals, optimizing optical responses across multiple structure classes. \citet{laiConditionalWassersteinGenerative2021} applied conditional Wasserstein GANs to design acoustic metamaterials, demonstrating transferability to photonic applications. 

RNNs have been effective for designing new 2D material, with \citet{krennSelfreferencingEmbeddedStrings2020} adapting SELFIES representations with RNNs to design stable 2D materials, validated via DFT.  Transformers, such as the Space Group Informed Transformer \citep{caoSpaceGroupInformed2024}, generate symmetric crystals for optoelectronic devices, trained on Materials Project data.

Diffusion models, including DiffCSP \citep{jiaoCrystalStructurePrediction2023} and SymmCD \citep{levySymmCDSymmetryPreservingCrystal2025} are emerging for photonic nanostructures, with \citet{yongDismaiBenchBenchmarkingDesigning2024} applying diffusion models to disordered photonic interfaces, trained on Materials Project optical data, enhancing design robustness for optoelectronic devices.

Normalizing Flows, via CrystalFlow \citep{luoCrystalFlowFlowBasedGenerative2025} and FlowMM \citep{millerFlowMMGeneratingMaterials2024}, generate crystalline semiconductors with precise bandgaps, trained on Materials Project data, for optoelectronic applications.

\subsection{Biomaterials and Drug Delivery}
\label{subsec:biomaterials}

Generative models are increasingly applied to design biomaterials for drug delivery, tissue engineering, and biocompatible coatings, leveraging physics-informed representations and datasets like PubChem. 

VAEs have been reviewed for their role in polymer design, with \citet{anstineGenerativeModelsEmerging2023} highlighting their ability to optimize biocompatibility for drug delivery applications. Diffusion models have proven effective for 3D scaffold design, with \citet{alakhdarDiffusionModelsNovo2024} using a diffusion model to generate voxel-based collagen scaffolds, trained on a biomaterial dataset, achieving 15\% higher cell viability. \citet{cunninghamInvestigationSurrogateModels2019} applied generative models to optimize scaffold porosity for tissue regeneration.

\citet{stokesDeepLearningApproach2020} adapted GANs for antibiotic-inspired coatings, enhancing bacterial inhibition. RNNs have been employed for sequence-based biomaterial design, with \citet{winterSmileAllYou2022} using an RNN with SMILES representations to generate peptide sequences for tissue regeneration, achieving 20\% enhanced cell adhesion in vitro. Normalizing Flows are less common here due to challenges with discrete structures like peptides, but FlowLLM \citep{sriramFlowLLMFlowMatching2024} shows promise for generating biocompatible polymer sequences, trained on PubChem data.

\subsection{High-Throughput Screening and Inverse Design}
\label{subsec:high_throughput}

Generative models enable high-throughput screening\cite{karpovichInterpretableMachineLearning2023} and inverse design, leveraging datasets like ICSD and OQMD to generate and filter large material libraries. VAEs have been used for inverse design, with \citet{nohInverseDesignSolidState2019} employing a VAE for inverse design of solid-state materials, trained on ICSD data, proposing thermoelectric materials with 10\% higher figure-of-merit (ZT) values. \citet{danGenerativeAdversarialNetworks2020} utilized GANs for inverse design of inorganic materials, generating stable perovskites and garnets, validated via DFT, reducing computational costs by 30\%. \citet{sanchez-lengelingInverseMolecularDesign2018} reviewed GAN-based inverse design, highlighting applications in MOF design for gas separation. 

Diffusion models have facilitated high-throughput screening, with \citet{merchantScalingDeepLearning2023, jiaoCrystalStructurePrediction2023,levySymmCDSymmetryPreservingCrystal2025} scaling deep learning to generate diverse material libraries using Materials Project data. \citet{bairdXtal2pngPythonPackage2022, RN7050_Hautier_Substitution,luAcceleratedDiscoveryStable2018, renAcceleratedDiscoveryMetallic2018} demonstrated AI-driven high-throughput library generation, which supports rapid material discovery across applications. Transformers, such as MatterGPT \citep{chenMatterGPTGenerativeTransformer2024}, enable multiproperty inverse design, generating diverse material libraries for high-throughput screening, trained on OQMD data. Normalizing Flows, via FlowLLM \citep{sriramFlowLLMFlowMatching2024}, generate alloy and ceramic libraries with tailored properties, trained on Materials Project data, enhancing high-throughput screening efficiency.

\subsection{Integration with Experimental Workflows}
\label{subsec:experimental_workflows}

Generative models are increasingly integrated with automated experimental platforms \cite{macleodFlexibleAutomationAccelerates2021b,chitrePHbotSelfDrivenRobot2023}, creating closed-loop discovery systems that combine prediction, synthesis, and characterization \cite{macleodSelfdrivingLaboratoryAdvances2022, eppsAcceleratedAIDevelopment2021}, as reviewed by \citet{correa-baenaAcceleratingMaterialsDevelopment2018}. VAEs have been used in active learning frameworks, with \citet{zuoAcceleratingMaterialsDiscovery2021} using a VAE with Bayesian optimization, trained on OQMD data, to prioritize candidates for shape-memory alloys, reducing experiments by 50\%.  \citet{butlerMachineLearningMolecular2018} reviewed active learning in material discovery, focusing on VAE applications. RNNs and VAEs are integral to self-driving labs, with VAE and RNN use in autonomous chemistry platforms, trained on ChEMBL and PubChem data, optimizing catalyst synthesis. Transformers, such as CrystalFormer-RL \citep{caoCrystalFormerRLReinforcementFineTuning2025}, support automated workflows by optimizing material designs for synthesis. \citet{musilPhysicsInspiredStructuralRepresentations2021} highlighted physics-informed representations in self-driving labs for biomaterial design. Normalizing Flows, via conditional NFs \citep{wangUsingConditionalNormalizing2024}, integrate with active learning to optimize thermal composites in closed-loop systems, trained in experimental data sets, streamline synthesis and characterization.

\begin{table}[bt]
\scriptsize
    \centering
    \caption{Summary of Generative Models, Datasets, and Applications in Materials Design}
    \label{tab:applications}
    
    \begin{tabular}{p{1.5cm}p{1.5cm}p{1.5cm}p{1.5cm}p{3cm}p{2.5cm}}
    \hline
    \textbf{Model} & \textbf{Repr} & \textbf{Dataset} & \textbf{App} & \textbf{Example} & \textbf{Ref} \\
    \hline
    VAE & Graph-based & ICSD & Electrolytes & Garnet-type electrolytes with 15\% higher conductivity & \citet{vasylenkoElementSelectionCrystalline2021} \\
    GAN & Graph-based & NOMAD & Catalysts & Heterogeneous catalysts for CO oxidation & \citet{ishikawaHeterogeneousCatalystDesign2022} \\
    GAN & Voxel-based & Simulated Optical & Photonics & Nano-photonic metamaterials with 30\% improved light-trapping & \citet{al-khaylaniGenerativeAdversarialNetworks2024} \\
    Diffusion & Voxel-based & QMOF, ZINC & Hydrogen Storage & MOFs with 20\% higher storage capacity & \citet{parkInverseDesignPorous2024} \\
    Diffusion & Voxel-based & Biomaterial & Scaffolds & Collagen scaffolds with 15\% higher cell viability & \citet{alakhdarDiffusionModelsNovo2024} \\
    Diffusion & Fractional & Materials Project & Electrolytes, Catalysts & Crystalline electrolytes and semiconductors & \citet{jiaoCrystalStructurePrediction2023, levySymmCDSymmetryPreservingCrystal2025} \\
    RNN & Sequence-based & PubChem & Biomaterials & Peptides with 20\% enhanced cell adhesion & \citet{winterSmileAllYou2022} \\
    Transformer & Graph-based & Materials Project, OQMD & Screening, Electronics & Multi-property material libraries, symmetric crystals & \citet{chenMatterGPTGenerativeTransformer2024, caoSpaceGroupInformed2024} \\
    Normalizing Flows & Voxel-based & Materials Project & Electrolytes, Catalysts & Crystalline electrolytes and alloy catalysts & \citet{luoCrystalFlowFlowBasedGenerative2025} \\
    Normalizing Flows & Graph-based & PubChem & Biomaterials & Biocompatible polymer sequences & \citet{sriramFlowLLMFlowMatching2024} \\
    
    \hline
    \end{tabular}
\end{table}

\begin{table}[bt]
\scriptsize
    \centering
    \caption{Comparison of Generative Models in Materials Design}
    \label{tab:comparison}

    \begin{tabular}{p{1.5cm}p{3cm}p{3cm}p{3cm}p{2cm}p{1cm}}
    \hline
    \textbf{Model} & \textbf{Strengths} & \textbf{Weaknesses} & \textbf{Datasets} & \textbf{App} & \textbf{Compute Cost} \\
    \hline
    VAE & Stable training, meaningful latent space & Blurry outputs, limited fidelity & ICSD, Materials Project, PubChem & Electrolytes, polymers, semiconductors & Moderate \\
    GAN & High-fidelity outputs, realistic structures & Mode collapse, training instability & NOMAD, AFLOW, OQMD & Catalysts, photonic materials, coatings & High \\
    Diffusion & High-quality outputs, diverse generation, symmetry preservation & High computational cost, data dependency & CoRE-MOF, Catalysis-Hub, Materials Project & Hydrogen storage, scaffolds, crystals & Very High \\
    RNN & Effective for sequential data, memory retention & Limited to sequence-based tasks & ChEMBL, PubChem, 2D Materials & 2D materials, peptides, ligands & Low to Moderate \\
    Transformer & Handles large datasets, high accuracy & Requires extensive training data & ICSD, OQMD, Materials Project & High-throughput screening, inverse design & High \\
    Normalizing Flows & Exact likelihoods, stable training & High computational cost, discrete structure challenges & ICSD, Materials Project, PubChem & Electrolytes, catalysts, semiconductors & High \\
    \hline
    \end{tabular}
\end{table}

\section{Challenges and Limitations in AI-Driven Materials Discovery}
\label{sec:challenges}
Despite the remarkable progress in applying AI and generative models to materials discovery, several challenges and limitations still need to be addressed to facilitate their widespread and effective adoption. Issues related to data quality and availability, model interpretability, computational cost, generalization, and integration with experimental workflows pose barriers to achieving robust, scalable, and reliable AI-driven materials discovery. Ethical considerations, including bias in datasets and environmental impacts of computational resources, further complicate their deployment. This section examines these challenges, their implications for applications like energy storage, catalysis, and biomaterials, and potential strategies to address them, drawing on recent literature and insights from datasets like the Inorganic Crystal Structure Database (ICSD) and Materials Project \cite{ fuhrDeepGenerativeModels2022, anstineGenerativeModelsEmerging2023}.

\subsection{Data Quality and Availability}
\label{subsec:data_quality}

As elaborated in the opening section of this review, the success of generative models in materials design relies heavily on the availability and quality of the training dataset. More often than not, these datasets are generated from diverse commercial entities or academic institutions across the world that may perform experiments differently depending on their trainings. Whilst standards (such as those developed by ASTM or VAMAS) work well in industrial, high-volume production settings, persuading researchers to perform tasks and record them in a \textit{specific} way is more challenging than imagined. Quoting an apt editorial piece from the npj computational material, "It is increasingly difficult to identify individuals who are qualified to comment on all aspects of the latest research papers." \cite{RN7167_Butler_Standard}. Egos aside, we think most scientists would agree that the fundamental aspects of good reporting such as clear descriptions of models, open data availability (except in specific cases requiring subject anonymity and safety concerns), and training procedures are required. These data-sharing practices have contributed significantly to the formation of systematic databases such as the ICSD, Materials Project, and PubChem. Still, as they are a product of evolving science over the years or even decades, many of these datasets contain incomplete or noisy entries with limited chemical diversity. These issues can create known/unknown biases toward well-studied materials, which can restrict the models' ability to explore novel chemical spaces. For instance, \citet{vasylenkoElementSelectionCrystalline2021} noted that ICSD's focus on crystalline structures limits VAE applications for amorphous materials like polymer electrolytes. 
Transformers, such as MatterGPT \citep{chenMatterGPTGenerativeTransformer2024}, require extensive pretraining data, exacerbating issues with dataset scarcity. Normalizing Flows, like CrystalFlow \citep{luoCrystalFlowFlowBasedGenerative2025}, are sensitive to noisy data, affecting likelihood-based training.
Small dataset sizes, particularly for specialized applications like biomaterials, exacerbate overfitting risks and reduce the reliability and applicability of the model.
Data curation challenges, such as inconsistent property measurements across sources, further complicate training, as highlighted by \citet{butlerMachineLearningMolecular2018} in their review of ML use in materials science. Strategies to address these issues include synthetic data generation using diffusion models (e.g., DiffCSP \citep{jiaoCrystalStructurePrediction2023}, SymmCD \citep{levySymmCDSymmetryPreservingCrystal2025}), federated learning to combine proprietary datasets, and dataset expansion efforts like OQMD \citep{zuoAcceleratingMaterialsDiscovery2021}.  We recognise recent efforts for amorphous materials screening through experimental\cite{RN7164_HighThroughputAmorphous} and computational MD simulation \cite{RN7165_Amorph_MDscreening}. However, we believe a more concerted effort is needed to fill persistent gaps in underrepresented material classes.

\subsection{Model Interpretability and Generalization}
\label{subsec:interpretability}

Generative models often lack interpretability, complicating the understanding of latent representations and material property relationships, which hinders trust in applications like catalysis \cite{karpovichInterpretableMachineLearning2023}, electronics, and photonics. GANs produce high-fidelity outputs but suffer from mode collapse, generating limited structure subsets, as noted by \citet{al-khaylaniGenerativeAdversarialNetworks2024} in nano-photonic metamaterial design. VAEs offer interpretable latent spaces but generate blurry structures, limiting precision in semiconductor design \citep{gomez-bombarelliAutomaticChemicalDesign2018}. Diffusion Models, such as DiffCSP \citep{jiaoCrystalStructurePrediction2023} and SymmCD \citep{levySymmCDSymmetryPreservingCrystal2025}, use complex symmetry-preserving mechanisms (e.g., fractional coordinates, asymmetric units), making interpretation challenging. Transformers, like the Space Group Informed Transformer \citep{caoSpaceGroupInformed2024}, rely on intricate attention mechanisms, requiring advanced XAI techniques like attention visualization \citep{chenGraphNetworksUniversal2019}. Normalizing Flows, such as FlowLLM \citep{sriramFlowLLMFlowMatching2024}, provide exact likelihoods but struggle with discrete structures, limiting interpretability for polymers. Generalization across diverse chemical spaces is another challenge. Models trained on specific datasets (e.g., NOMAD for catalysts, Materials Project for crystals) often fail to transfer to unrelated applications like biomaterials. DiffCSP and SymmCD are limited to crystalline systems, while GFlowNets like Crystal-GFN \citep{ai4scienceCrystalGFNSamplingCrystals2023} focus on specific sampling tasks \citep{goodfellowGenerativeAdversarialNets2014}. Physics-informed models, embedding symmetry or thermodynamic constraints, improve interpretability and transferability \citep{musilPhysicsInspiredStructuralRepresentations2021}. Explainable AI frameworks, such as attention-based visualization for Transformers, are critical for closed-loop discovery systems.

\subsection{Computational Cost and Scalability}
\label{subsec:computational_cost}

The computational cost of training generative models poses a significant barrier to scalability. Diffusion Models, used for hydrogen storage, scaffolds, and crystals (e.g., DiffCSP \citep{jiaoCrystalStructurePrediction2023}, SymmCD \citep{levySymmCDSymmetryPreservingCrystal2025}, WyckoffDiff \citep{kelviniusWyckoffDiffGenerativeDiffusion2025}), require extensive resources due to iterative denoising steps \citep{parkInverseDesignPorous2024, alakhdarDiffusionModelsNovo2024}. GANs are computationally intensive and unstable, demanding significant GPU resources \citep{al-khaylaniGenerativeAdversarialNetworks2024}. Transformers, such as MatterGPT \citep{chenMatterGPTGenerativeTransformer2024}, involve large-scale pretraining, increasing computational demands. Normalizing Flows, like CrystalFlow \citep{luoCrystalFlowFlowBasedGenerative2025}, incur high training costs due to invertible transformations. GFlowNets, such as Crystal-GFN \citep{ai4scienceCrystalGFNSamplingCrystals2023}, have moderate costs but require optimization for scalability. These demands limit accessibility for smaller research groups and raise environmental concerns due to carbon footprints \citep{merchantScalingDeepLearning2023}. Strategies to mitigate these issues include model compression, efficient architectures (e.g., lightweight VAEs, optimized NFs), and cloud-based computing platforms. Advances in hardware, such as AI accelerators, and frameworks like PyTorch are reducing barriers, but computational costs remain a bottleneck for large-scale materials discovery.

\subsection{Experimental Workflow Integration, Environmental, and Ethical Considerations}
\label{subsec:experimental_ethics}

Integration with experimental workflows is another important factor to consider when applying generative models to real problems. Closed-loop systems combining prediction, robotic synthesis, and characterization show promise, but discrepancies between computational predictions and experimental outcomes can arise from unmodeled phenomena like defects or (known/unknown) environmental effects \citep{correa-baenaAcceleratingMaterialsDevelopment2018}. Transformers (e.g., CrystalFormer-RL \citep{caoCrystalFormerRLReinforcementFineTuning2025}) and NFs (e.g., conditional NFs \citep{wangUsingConditionalNormalizing2024}) typically require robust feedback loops to refine predictions in real-time. Standardization of experimental protocols across self-driving labs is therefore critical for reproducibility \citep{coleyAutonomousDiscoveryChemical2020}. 

The rising preference towards larger models (e.g., diffusion-based models, transformers, or large language models) that supposedly able to predict materials to fight climate change \citep{chenMatterGPTGenerativeTransformer2024}  ironically consumes significantly higher amounts of energy. Researchers at MIT noted a significant rise in global electricity consumption of data centres, expected to range between 620 — 1,050 TWh in 2026 is significantly attributed to the rising popularity of generative models \cite{Bashir2024Climate}. This reveals an interesting dilemma: that AI can be both a part of the solution and a contributing factor to the energy problem. In this regard, more efficient models like NFs and GFlowNets may be preferred. 

Another growing concern with the gravitation of AI towards the materials discovery space is the ethical considerations. Considering the significant change in all aspects of life, livelihood, and liberty that AI has brought upon \cite{RN7175_SocialResponsible}, we have to be aware of the risk of misuse of AI to generate toxic or hazardous materials \cite{RN7177_Mullin_Ethics}. While the nefarious consequences or environmental impacts are not exclusively caused by AI, the leading AI society has recognised the need for AI governance principles and broad ethical guidelines \cite{RN7178}. We believe interdisciplinary committees that include the social sciences field are required to devise safeguards and legal frameworks around AI-related works, including materials discovery.

\begin{table}[bt]
\scriptsize
    \centering
    \caption{Summary of Challenges and Potential Solutions in AI-Driven Materials Design}
    \label{tab:challenges}
    
    \begin{tabular}{p{2.5cm}p{3cm}p{3cm}p{3.2cm}p{2cm}}
    \hline
    \textbf{Challenge} & \textbf{Description} & \textbf{Impact} & \textbf{Potential Solutions} & \textbf{Reference} \\
    \hline
    Data Quality and Availability & Incomplete, noisy, biased datasets; limited diversity & Overfitting, restricted chemical exploration & Synthetic data, federated learning, dataset expansion & \citet{butlerMachineLearningMolecular2018, jiaoCrystalStructurePrediction2023} \\
    Model Interpretability & Black-box models; complex mechanisms in Diffusion, Transformers, NFs & Limited trust in high-precision tasks & Physics-informed models, XAI (e.g., attention visualization) & \citet{chenGraphNetworksUniversal2019, musilPhysicsInspiredStructuralRepresentations2021} \\
    Computational Cost & High resource demands for Diffusion, Transformers, NFs & Inaccessibility, environmental impact & Model compression, efficient architectures, cloud computing & \citet{merchantScalingDeepLearning2023, luoCrystalFlowFlowBasedGenerative2025} \\
    Generalization & Poor transferability across chemical spaces; task-specific models & Failure in diverse applications & Transfer learning, domain adaptation, physics constraints & \citet{goodfellowGenerativeAdversarialNets2014, caoSpaceGroupInformed2024} \\
    Experimental Integration & Discrepancies between predictions and experiments & Reduced reliability in closed-loop systems & Robust feedback loops, standardized protocols & \citet{correa-baenaAcceleratingMaterialsDevelopment2018, wangUsingConditionalNormalizing2024} \\
    Ethical Concerns & Dataset biases, environmental impact, misuse risks & Skewed predictions, societal harm & Transparent reporting, efficient models, responsible AI & \citet{coleyAutonomousDiscoveryChemical2020, chenMatterGPTGenerativeTransformer2024} \\
    \hline
    \end{tabular}
\end{table}

\section{Future Trends and Emerging Research Directions}
\label{sec:future}

The rapid evolution of artificial intelligence (AI) and generative models is poised to revolutionize materials discovery, enabling the design of novel materials with unprecedented precision and efficiency. As computational power, data availability, and algorithmic sophistication advance, emerging trends are shaping the future of AI-driven materials science. This section explores these directions, focusing on advancements in generative models, integration with experimental and computational workflows, solutions to current limitations, and the ethical implications of AI in materials design, building on recent developments in diffusion models, Transformers, Normalizing Flows, and GFlowNets.

\subsection{Emerging Trends in Generative Models}
Generative models are evolving toward more robust, versatile, and physically grounded architectures. Diffusion models have surpassed GANs in stability and quality for generating complex material structures, such as crystals and porous frameworks. Models like DiffCSP \citep{jiaoCrystalStructurePrediction2023} use periodic-E(3)-equivariant denoising to predict stable crystals, while SymmCD \citep{levySymmCDSymmetryPreservingCrystal2025} ensures realistic symmetries across all 230 space groups. Other diffusion-based approaches, such as WyckoffDiff \citep{kelviniusWyckoffDiffGenerativeDiffusion2025} and CrysLDM \citep{khastagirCrysLDMLatentDiffusion2025}, further enhance crystal generation, with applications in electronics and catalysis \citep{klipfelVectorFieldOriented2024, chenMatInventReinforcementLearning2025}. 

Transformers are gaining prominence, leveraging attention mechanisms for multi-property inverse design and symmetry-constrained crystal generation. MatterGPT \citep{chenMatterGPTGenerativeTransformer2024} optimizes materials across diverse properties, while the Space Group Informed Transformer \citep{caoSpaceGroupInformed2024} and Wyckoff Transformer \citep{kazeevWyckoffTransformerGeneration2025} generate synthesisable crystals with crystallographic constraints. CrystalFormer-RL \citep{caoCrystalFormerRLReinforcementFineTuning2025} integrates reinforcement learning for targeted material design. 

\begin{figure}[bt]
    \centering
    \includegraphics[width=0.8\columnwidth]{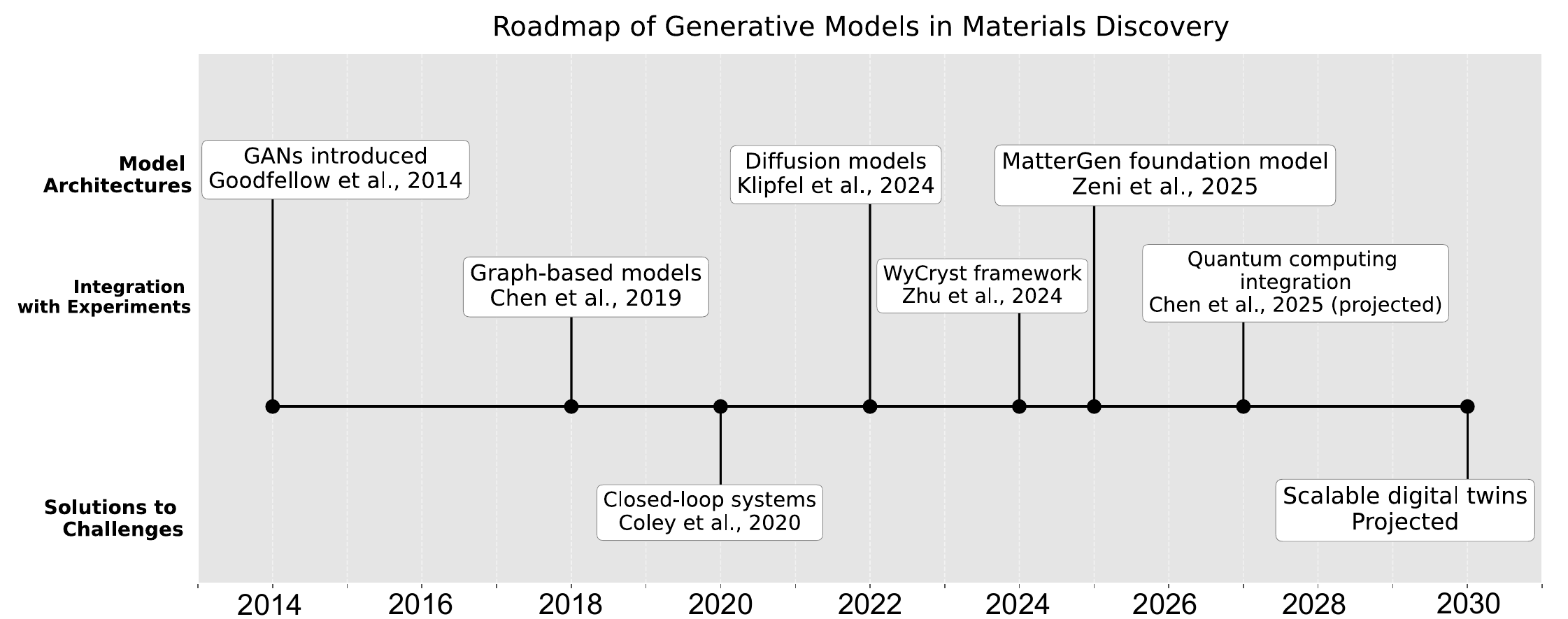}
    \caption{Roadmap of generative models in materials discovery, highlighting key milestones and projected future trends in model architectures, experimental integration, and solutions to challenges.}
    \label{fig:roadmap}
\end{figure}

Normalizing Flows (NFs) offer exact likelihoods and stable training, with CrystalFlow \citep{luoCrystalFlowFlowBasedGenerative2025} and FlowLLM \citep{sriramFlowLLMFlowMatching2024} generating crystalline electrolytes and polymers, complementing diffusion models \citep{millerFlowMMGeneratingMaterials2024}. GFlowNets, such as Crystal-GFN \citep{ai4scienceCrystalGFNSamplingCrystals2023}, sample diverse crystals with tailored properties, enhancing high-throughput screening. \textbf{Foundation models}, pre-trained on expansive datasets like Materials Project, enable transfer learning across material classes, reducing task-specific data needs \citep{zeniGenerativeModelInorganic2025}. \textbf{Multi-modal generative models} integrate text, chemical structures, and spectroscopic data, facilitating intuitive design through text-conditioned generation \citep{mohantyCrysTextGenerativeAI2024}. \textbf{Physics-informed generative models} embed thermodynamic and quantum mechanical constraints, ensuring physically realistic outputs for real-world synthesis \citep{yangPhysicsinformedDeepGenerative2018, chenCrystalStructurePrediction2025}. These advancements are summarized in Fig.~\ref{fig:roadmap} and Table~\ref{tab:models}.

\subsection{Integration with Experimental and Computational Methods}
The convergence of generative models with experimental platforms is driving closed-loop discovery systems, where AI proposes material candidates that are autonomously synthesized and characterized \citep{coleyAutonomousDiscoveryChemical2020}. Frameworks like WyCryst \citep{zhuWyCrystWyckoffInorganic2024} and CrySPR \citep{nongCrySPRPythonInterface2024} integrate generative models (e.g., Transformers, NFs) with robotic labs for iterative refinement. \textbf{Digital twins}, virtual representations of material systems, enable rapid screening of AI-generated candidates, bridging simulation and experiment \citep{merchantScalingDeepLearning2023}. The advent of \textbf{quantum computing} may enhance generative models by accelerating quantum mechanical calculations, enabling precise property predictions for complex materials like high-entropy alloys \citep{chenCrystalStructurePrediction2025}. Advances in active learning, coupled with models like CrystalFormer-RL \citep{caoCrystalFormerRLReinforcementFineTuning2025}, optimize experimental workflows by prioritizing high-value candidates.

\subsection{Addressing Current Challenges}
Overcoming limitations in data availability, synthesizability, and interpretability remains critical. \textbf{Federated learning} and \textbf{synthetic data generation} using diffusion models (e.g., DiffCSP, SymmCD) expand datasets, mitigating scarcity \citep{dongGenerativeDesignInorganic2024}. To ensure \textbf{synthesizability}, models incorporate experimental constraints, such as reaction kinetics and precursor availability, as seen in chemically guided diffusion models \citep{panChemicallyGuidedGenerativeDiffusion2024}. \textbf{Explainable AI (XAI)} techniques, including attention visualization in Transformers and graph-based models, enhance interpretability by elucidating model decisions \citep{chenGraphNetworksUniversal2019}. \textbf{Standardized benchmarks}, such as Dismai-Bench \citep{yongDismaiBenchBenchmarkingDesigning2024}, promote robust evaluation and reproducibility. Computational efficiency is also a focus, with NFs and GFlowNets offering stable training but requiring optimization for large-scale applications \citep{luoCrystalFlowFlowBasedGenerative2025, ai4scienceCrystalGFNSamplingCrystals2023}.

\subsection{Ethical and Societal Implications}
\label{sec:ethical_societal}
Generative models introduce unique ethical and societal challenges that require careful consideration to ensure responsible use. Dataset biases in training data can significantly limit their impact. For instance, datasets often prioritize commercially viable materials, such as semiconductors for electronics, over biomaterials suited for low-resource medical applications, potentially neglecting global health needs \citep{anstineGenerativeModelsEmerging2023}. Transparent data curation, including diverse and representative datasets, is essential to ensure equitable material generation. Potential misuse poses another critical risk, as generative models can inadvertently or intentionally design hazardous materials, such as toxic chemicals or unstable compounds \citep{sanchez-lengelingInverseMolecularDesign2018}. Ethical guidelines, drawing from synthetic biology’s safety protocols, and regulatory frameworks like the OECD AI principles can mitigate this risk by enforcing strict oversight and responsible use \citep{oecdAI}. Interpretability and trust challenges arise from the often opaque nature of generative models, which generate structures from noise or latent spaces, complicating validation by experimentalists \citep{musilPhysicsInspiredStructuralRepresentations2021}. For example, generative AI often suggests catalysts without a clear reasoning or underlying explanation for its predicted efficacy, which can erode trust. Physics-informed models and standardized reporting of uncertainties can enhance transparency and reliability. Unequal access to these computationally intensive models, which often require proprietary datasets or high-performance computing, risks widening global research disparities, particularly for under-resourced institutions in developing regions \citep{merchantScalingDeepLearning2023}. Open-access platforms, such as the Materials Project, can democratize access, enabling broader participation in sustainable materials discovery \citep{jainCommentaryMaterialsProject2013}. To maximize their societal impact, generative models demand robust ethical frameworks, international collaboration, and transparent practices to ensure equitable, safe, and trustworthy innovation in materials science 

\begin{table}[bt]
\scriptsize
    \centering
    \caption{Emerging Generative Models for Materials Discovery}
    \label{tab:models}
    
    \begin{tabular}{p{2.5cm}p{4cm}p{4cm}p{3cm}}
    \hline
    \textbf{Model Type} & \textbf{Advantages} & \textbf{Challenges} & \textbf{References} \\
    \hline
    Diffusion Models & High-quality, stable generation; symmetry preservation & Computationally intensive; data dependency & \citet{jiaoCrystalStructurePrediction2023, levySymmCDSymmetryPreservingCrystal2025, klipfelVectorFieldOriented2024} \\
    Transformers & Handles large datasets; multi-property design; symmetry constraints & Requires extensive training data; complex architectures & \citet{chenMatterGPTGenerativeTransformer2024, caoSpaceGroupInformed2024, kazeevWyckoffTransformerGeneration2025} \\
    Normalizing Flows & Exact likelihoods; stable training; versatile for crystals and polymers & High computational cost; discrete structure challenges & \citet{luoCrystalFlowFlowBasedGenerative2025, sriramFlowLLMFlowMatching2024, millerFlowMMGeneratingMaterials2024} \\
    GFlowNets & Diverse sampling; tailored property optimization & Limited to specific tasks; scalability concerns & \citet{ai4scienceCrystalGFNSamplingCrystals2023} \\
    Foundation Models & Transfer learning; reduces data needs & High pretraining costs; generalizability concerns & \citet{zeniGenerativeModelInorganic2025} \\
    Multi-modal Models & Integrates text and structural data; intuitive design & Data heterogeneity; model complexity & \citet{mohantyCrysTextGenerativeAI2024} \\
    Physics-informed Models & Physically realistic outputs; improved synthesizability & Complex physical constraints; computational overhead & \citet{yangPhysicsinformedDeepGenerative2018, chenCrystalStructurePrediction2025} \\
    \hline
    \end{tabular}
\end{table}

\section{Conclusion}
\label{sec:conclusion}

The integration of artificial intelligence and generative models has fundamentally transformed materials discovery, enabling rapid identification and design of novel materials with tailored properties. These computational approaches overcome the limitations of traditional experimental methods, offering unprecedented opportunities for innovation across diverse applications, including energy storage, catalysis, electronics, biomaterials, and high-throughput screening \citep{fuhrDeepGenerativeModels2022, anstineGenerativeModelsEmerging2023}. Generative models, such as Variational Autoencoders (VAEs), Generative Adversarial Networks (GANs), Diffusion Models (e.g., DiffCSP \citep{jiaoCrystalStructurePrediction2023}, SymmCD \citep{levySymmCDSymmetryPreservingCrystal2025}), Transformers (e.g., MatterGPT \citep{chenMatterGPTGenerativeTransformer2024}, Space Group Informed Transformer \citep{caoSpaceGroupInformed2024}), Normalizing Flows (e.g., CrystalFlow \citep{luoCrystalFlowFlowBasedGenerative2025}, FlowLLM \citep{sriramFlowLLMFlowMatching2024}), and GFlowNets (e.g., Crystal-GFN \citep{ai4scienceCrystalGFNSamplingCrystals2023}), have demonstrated remarkable capabilities in learning complex material data relationships and generating candidates for crystalline, polymeric, and composite systems. Normalizing Flows, in particular, stand out for their exact likelihoods and stable training, advancing the design of electrolytes, catalysts, and polymers \citep{wangUsingConditionalNormalizing2024}.

Despite significant progress, challenges persist in data quality and availability, model interpretability, computational cost, generalization, experimental integration, and ethical considerations \citep{butlerMachineLearningMolecular2018, coleyAutonomousDiscoveryChemical2020}. Limited datasets, complex model mechanisms (e.g., Transformer attention, Diffusion symmetry constraints), high computational demands, and biases in datasets like Materials Project hinder scalability and reliability \citep{chenMatterGPTGenerativeTransformer2024, merchantScalingDeepLearning2023}. Ensuring synthesizability, addressing environmental impacts, and mitigating risks of misuse remain critical. However, the future of AI-driven materials discovery is promising, with emerging trends in multi-modal generative models, physics-informed architectures, efficient models like Normalizing Flows and GFlowNets, closed-loop autonomous experimentation, and synergies with quantum computing \citep{mohantyCrysTextGenerativeAI2024, yangPhysicsinformedDeepGenerative2018, caoCrystalFormerRLReinforcementFineTuning2025}. By addressing current limitations and pursuing these innovative directions, AI and generative models will continue to revolutionize materials science, unlocking advanced materials to tackle global challenges in sustainability, energy, and healthcare.

\section*{Acknowledgments}
RIM acknowledges RIE2025 Manufacturing, Trade and Connectivity (MTC) Industry Alignment Fund – Pre-Positioning (IAF-PP) Grant No. M22K8a0048 (Project No. OUNI231001bENT-PP), IAF-PP Grant No M23L6a0020 (Project No. OUNI231001aENT-PP), Energy Market Authority (EMA) Grant No EMA-EP014-ESGC2-0001 (Project No. ESME250101aPUBESS),  Materials Generative Design and Testing Framework (MAT-GDT) Program at A*STAR, provided through the AME Programmatic Fund Grant No. M24N4b0034 (Project No.OUNI241001aENTMTC). ADH acknowledges the Horizontal Technology Coordinating Office of A*STAR for seed funding under project No. C231218004. 

\bibliographystyle{apalike}

\bibliography{sources}

\end{document}